\documentclass[10pt]{article}
\usepackage{setspace}
\usepackage{amsmath,amsthm,comment}
\usepackage{algorithm}
\usepackage{algorithmic}
\usepackage{amsfonts}
\usepackage{graphicx}
\usepackage{multirow}
\usepackage{booktabs}
\usepackage[top=1in, bottom=1in, left=1in, right=1in]{geometry}
\usepackage{subfigure}
\usepackage{enumerate}
\usepackage{dsfont}
\usepackage{bm}
\usepackage{natbib}
\usepackage{xcolor}
\usepackage{color}
\usepackage{caption}
\usepackage[colorlinks=true,citecolor=blue,linkcolor=blue,urlcolor=blue]{hyperref}
\allowdisplaybreaks
\numberwithin{equation}{section}
\theoremstyle{definition}

\theoremstyle{remark}

\newcommand{\bD}{\boldsymbol{D}}
\newcommand{\bM}{\boldsymbol{M}}
\newcommand{\bx}{\boldsymbol{x}}

\newcommand{\bN}{\boldsymbol{N}}

\newcommand{\bZ}{\boldsymbol{Z}}
\newcommand{\bY}{\boldsymbol{Y}}
\newcommand{\bW}{\boldsymbol{W}}
\newcommand{\bw}{\boldsymbol{w}}

\newcommand{\balpha}{\boldsymbol{\alpha}}
\newcommand{\bbeta}{\boldsymbol{\beta}}
\newcommand{\bgamma}{\boldsymbol{\gamma}}
\newcommand{\btheta}{\boldsymbol{\theta}}

\newcommand{\bPsi}{\boldsymbol{\Psi}}
\newcommand{\bpsi}{\boldsymbol{\psi}}
\newcommand{\bPhi}{\boldsymbol{\Phi}}
\newcommand{\bphi}{\boldsymbol{\phi}}

\begin{document}
%\title{Individual claim reserving using claim description }
\title{Combining Structural and Unstructured Data: A Topic-based Finite Mixture Model for Insurance Claim Prediction}
\author{
	Yanxi Hou\footnote{\scriptsize School of Data Science, Fudan University. Email: {yxhou@fudan.edu.cn}},\quad
    Xiaolan Xia\footnote{\scriptsize School of Data Science, Fudan University. Email: {21210980119@m.fudan.edu.cn}},\quad
	Guangyuan Gao\footnote{\scriptsize Center for Applied Statistics and School of Statistics, Renmin University of China. Corresponding author, Email: {guangyuan.gao@ruc.edu.cn}}
}
\date{}
\maketitle
{\begin{abstract}

Modeling insurance claim amounts and classifying claims into different risk levels are critical yet challenging tasks. 
Traditional predictive models for insurance claims often overlook the valuable information embedded in claim descriptions. 
This paper introduces a novel approach by developing a joint mixture model that integrates both claim descriptions and claim amounts. 
Our method establishes a probabilistic link between textual descriptions and loss amounts, enhancing the accuracy of claims clustering and prediction. 
In our proposed model, the latent topic/component indicator serves as a proxy for both the thematic content of the claim description and the component of loss distributions. Specifically, conditioned on the topic/component indicator, the claim description follows a multinomial distribution, while the claim amount follows a component loss distribution. We propose two methods for model calibration: an EM algorithm for maximum a posteriori estimates, and an MH-within-Gibbs sampler algorithm for the posterior distribution. The empirical study demonstrates that the proposed methods work effectively, providing interpretable claims clustering and prediction.

\end{abstract}}

{\em Keywords:  Insurance claim prediction; Text analytics; Finite mixture model; Dirichlet multinomial mixtures; MCMC}

\section{Introduction}

Modeling insurance loss data and classifying losses coming from different sources are important yet challenging tasks.
It is well known that the insurance losses coming from different sources are heterogeneous as reflected in multimodality, skewness, and heavy tail of the loss distributions. 
These features are difficult to deal with in actuarial practice. 
One appealing approach is to model the insurance losses using $K$-component finite mixture models, which  capture the heterogeneity in the data and further allow for the mixture components to represent groups in the population.
These finite mixture models usually focus on modeling the loss severity itself, 
but less studies are made to explore the  causes for different components in the mixture model. 
%{It has been already noted that given different risks assigned to each of the groups, augmenting the finite mixture model with a concomitant model for the weights  would allow classifying observations into these groups and thus enable an improved risk evaluation \citep{DaytonC.Mitchell1988CLM}.}

Different types of mixture models have been studied in the literature. 
\citet{1999Bayesian} proposed modeling losses with a mixture of exponential distributions with the maximum likelihood (ML) estimation based on the Newton‘s algorithm. 
While this model is useful in some actuarial applications, the mode of this model is at zero and the distribution is completely monotonic, which may result in a poor fit in the case of modeling heavy-tail losses \citep{WangJunfeng2006Agmf}.
\citet{2006Toward} tried to address this issue by proposing a flexible mixture model that  includes not only exponential components but also gamma, log-normal and Pareto components with the restriction that either weight associated with gamma or log-normal component equals zero. 
\citet{2010Modeling} proposed modeling and evaluating insurance losses via mixtures of Erlang distributions with the EM algorithm for parameter estimation.
\citet{2016Modeling} extended the mixture modeling beyond the component of  Erlang family and did not impose any restriction on the parameters. 
{The four-parameter distribution family, the generalized beta type-II (GB2), also known as the transformed beta distribution, has been proposed for modeling insurance losses \citep{ChanJ.S.K.2018MILU}. }
 \citet{2019On} modeled left-truncated loss insurance data using a non-Gaussian finite mixture model based on any combination of the gamma, log-normal and Weibull distributions with the EM algorithm for parameter estimation. 

Most of the literature applies the finite mixture models to  the numerical  loss data. 
However, in actuarial practice, there are usually non-numerical data available, for example, short text of claim description.
When it comes to the unstructured data, the traditional modeling methods are not suitable for these types of data, and thus the information hidden behind them is hard to incorporate into the model. 
In the literature of machine learning research, 
one common approach to deal with text data is topic model.
Early research on short text topic model mainly focused on using external information to enrich text features, including  \citet{2006A,2008Learning,2011Transferring}.
%\cite{2006A} proposed a short text similarity measurement method based on search fragments; 
 %\cite{2008Learning} learned hidden themes from a large number of external resources, thereby enriching short texts.
%\cite{2011Transferring} learned short text topics through transfer learning in assisted long text data. 
These methods are useful in certain specific fields, 
but not widely used due to the difficulty in obtaining useful external datasets. 
Another way is to add strong assumptions, such as assuming that each article contains only one topic, or assuming that the topic distribution of each article in the entire corpus is the same. 
%\blue{Zhao used a mixture of unigrams model to analyze Twitter data.}
Dirichlet Multinomial Mixture (DMM) model is a probabilistic generative model for short length documents. 
Compared with LDA \citep{Blei-2003}, this simplified text generation process can effectively solve the sparsity problem of short texts.	
\citet{1999Using} introduced an algorithm for learning unlabeled documents  for the DMM model.
\citet{2014A} propose a collapsed Gibbs sampling algorithm for the DMM model for short text clustering. 
Our work integrates a finite mixture model for claim loss with a DMM for claim description.

In this paper, we propose a novel approach to build a topic-based finite mixture model to jointly model the claims amount and the claim description text.
This approach  establishes a bridge between the loss amount and claim causes from a probabilistic generative perspective. 
The novelty of our method is to use claims description to better identify the degree of severity of each claim.
%that is, a detailed claim cause learned through the model. 
%Moreover, the risk prediction can be made based on the description text data without knowing the exact component the loss come from, which is of practical interests in actuarial practice. 
{Our proposed methods provide an interpretation of the components in the finite mixture model for claims amount.
More specifically, the proposed model provides a topic-based decomposition of the loss distributions, 
which is an insight into the causes of heterogeneity like multimodality, skewness, and heavy tail phenomenons.}
%\blue{To uniformly model loss data and text data, our approach is developed with components from Dirichlet multinomial distributions at the text level and parametric, non-Gaussian families of distributions used in actuarial modeling at the loss level.}
%We construct one basic model and one extended models based on the choice of loss distribution type and whether to include the prior of the loss distribution or not. 
We propose two parameter estimation methods:  the EM algorithm for the maximum a posteriori (MAP) estimate and the Gibbs sampling for the posterior distribution. %Model selection is possible by using information criteria, and the fitted models can be used to estimate risk measures for the data, such as VaR and TVaR, for practical purpose. 
%Our data analysis results show that the proposed methods work well. Our work provides new valuable tools in the area of insurance loss modeling, claim classification and risk evaluation

The rest of this paper is organized as follows. In Section 2, we propose a loss Dirichlet multinomial mixture model in the Bayesian framework. We also propose an EM algorithm and a Gibbs sampling algorithm  for the Bayesian model inference. 
In Section 3, we illustrate the proposed methods by analyzing a claims dataset.
Section 4 concludes with important findings.

\section{Loss Dirichlet multinomial mixture model}\label{sec:LDMM}

A prototype of claim case in our method is a triplet $(Y,\bD, Z)$, which contains two observed information, 
the claim loss denoted as $Y$, 
the claim description text  denoted as $\bD$, 
and the unobserved categorical variable denoted as $Z$.
A loss $Y$ is assumed to follow a finite mixture model \citep{2000Finite}, where the unobserved categorical variable $Z$ is the component indicator variable.
A document $\bD$ contains a collection of words $\langle D_{1}, D_{2}, D_3, \cdots, D_{|D|}\rangle$,
where $D_{j}$ is the $j$th word in the document and $|D|$ is the length of the document.
The words $D_j, j=1,\ldots, |D|$ can take values from a vocabulary $V=\{ w_1,w_2,\cdots, w_{|V|}\}$, i.e., $D_j\in V$. 
The words $D_j$ in a  document are assumed to be generated from a Dirichlet multinomial mixture (DMM) model \citep{1999Using}, 
i.e., the topic $Z$ of a document follows a discrete distribution and given the topic the words $D_j$  follow another discrete distribution.
The DMM model assumes that  the parameters in both discrete distributions follow a Dirichlet prior, respectively. 
Moreover, we make a standard naive Bayes assumption that the words of a document are generated independently of context, that is, independently of the other words in the same document given the topic $Z$, 
and we also assume that the document length $|D|$ is independent of the topic as well as the words.
We call this proposed model as a {\it Loss Dirichlet Multinomial Mixture (LDMM)}  model.
A key feature of the LDMM model is that  the finite mixture model of claim loss and the Dirichlet multinomial mixture model of claim description share the same component/topic indicator (latent) variable $Z$,
and given  the categorical variable $Z$, the loss $Y$ and the claim description text $\bD$ are assumed to be  independent.

We establish the LDMM model in a Bayesian network framework as follows:

\begin{equation}\label{LDMM}
	\begin{aligned}
		Z_i|\boldsymbol{\theta} ~ \overset{i.i.d.}{\sim}& ~ Dis(\boldsymbol\theta), i=1,\ldots,n,\\
				Y_i|Z_i, \boldsymbol\Phi ~\overset{i.i.d.}{\sim}& ~ p_{Z_i}(\boldsymbol\phi_{Z_i}), ~ i=1,\ldots,n, \\
		D_{i,j}|Z_i, \boldsymbol\Psi ~\overset{i.i.d.}{\sim}& ~ Dis(\boldsymbol\psi_{Z_i}), i=1,\ldots,n, j=1,\ldots,|\bD_i|, \\
		\boldsymbol{\theta}~{\sim}& ~ Dir(\boldsymbol{\alpha}), \\
	\boldsymbol{\phi}_k ~{\sim}& ~ q_k(\boldsymbol{\beta}_k), k=1,\ldots,K,\\
		\boldsymbol{\psi}_k~\overset{i.i.d.}{\sim}& ~ Dir(\boldsymbol{\gamma}), k=1,\ldots, K, 
	\end{aligned}
\end{equation}
where $Dis$ stands for the discrete distribution, $Dir$ stands for the Dirichlet distribution, $Z_i\in\{1,\ldots,K\}$, $Y_i\in(0,\infty)$, $D_{i,j}\in V$, $\boldsymbol{\Phi}=\{\boldsymbol{\phi}_1,\ldots,\boldsymbol{\phi}_K\}$, $\boldsymbol{\Psi}=\{\boldsymbol{\psi}_1,\ldots,\boldsymbol{\psi}_K\}$,  $p_k$ is the $k$th component loss distribution, $|\bD_i|$ is the length of the $i$th document and the last three lines specify the priors with the hyper-parameters $\balpha,\bbeta_1,\ldots,\bbeta_K,\boldsymbol\gamma$.
Note that $\boldsymbol{\theta}, \boldsymbol\alpha$ are $K$-dimensional vectors, and $\boldsymbol{\psi}_1, \ldots, \boldsymbol{\psi}_K, \boldsymbol{\gamma}$ are $|V|$-dimensional vectors.

\begin{figure}[h!]
	\centering
	\includegraphics[width=0.5\linewidth]{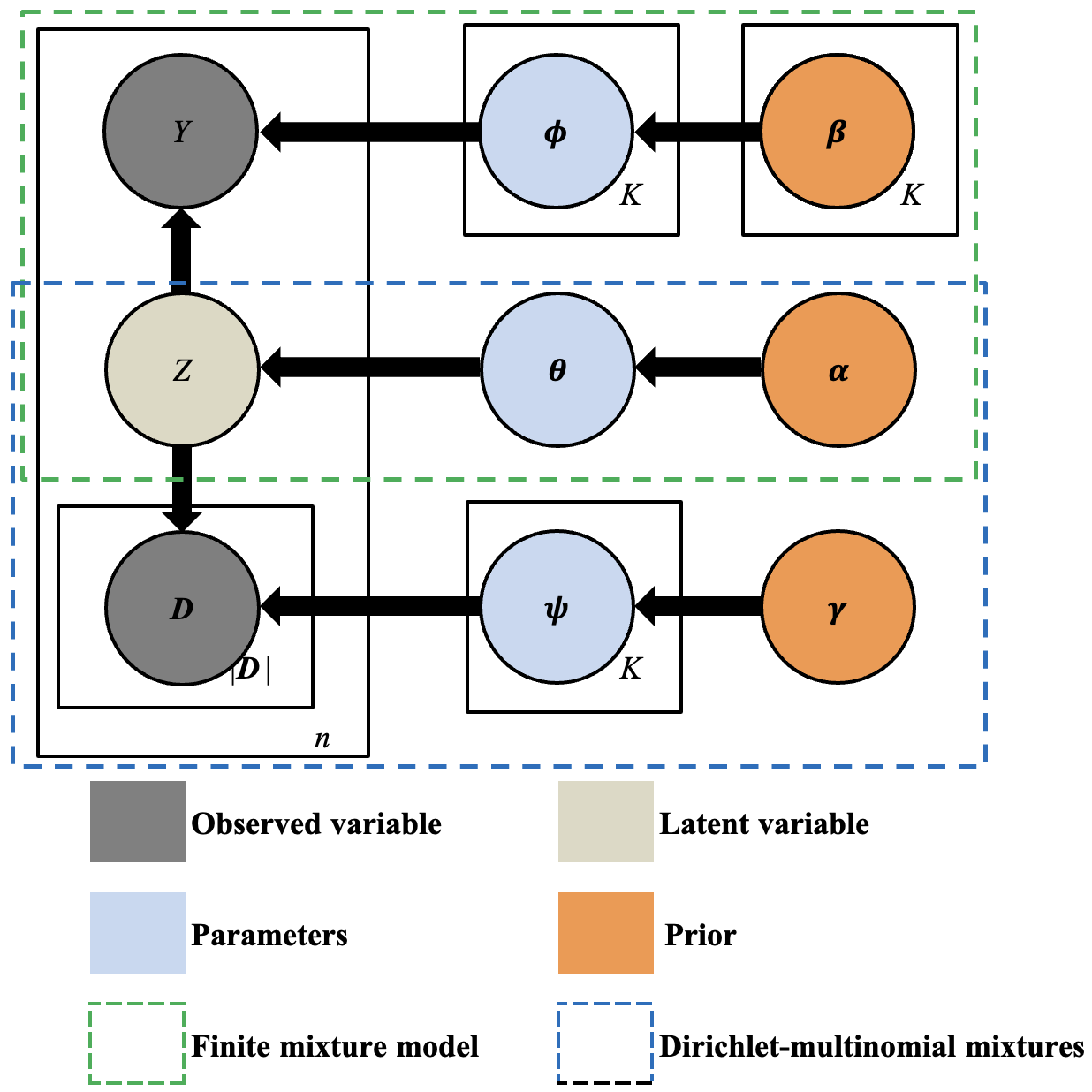}
	\caption{Graphical illustration of the LDMM model}	\label{fig:LDMM}
\end{figure}

Figure \ref{fig:LDMM} sketches a graph of the proposed LDMM model. 
The Bayesian network framework has two advantages over a frequentist framework: 
first, the prior of $\boldsymbol{\psi}$ introduces a smoothing effect which can prevent zero probabilities for infrequently occurring words;
{second, the Bayesian MCMC methods can simulate the posterior predictive distribution of claim amounts which is essential for risk management.}  

\subsection{Relevant distributions in the LDMM model}

Now we derive several joint/marginal/conditional distributions which will be used in the EM algorithm for the maximum a posteriori (MAP) estimate and the Gibbs sampler for the posterior distribution of the parameters and latent variable.
The posterior distribution of $\bPhi,\btheta,\bPsi,\bZ$ given the data $\bY,\bD$ is as follows:
\begin{equation}\label{posterior}
\begin{aligned}
	p(\bPhi,\btheta,\bPsi,\bZ|\bY,\bD)&\propto p(\bY,\bD|\bPhi,\btheta,\bPsi,\bZ)p(\bPhi,\btheta,\bPsi,\bZ) \\
	&= p(\bY|\bPhi,\bZ)p(\bD|\bPsi,\bZ)p(\bZ|\btheta)p(\btheta)p(\bPhi)p(\bPsi).
\end{aligned}
\end{equation}
Note a slight abuse of notation $\bD$. At the beginning of Section \ref{sec:LDMM}, $\bD$ denotes a general document containing several words, while here $\bD$ denotes a sample of observations $\bD=\{\bD_1,\ldots,\bD_n\}$. Whenever a misunderstanding might arise, we will clarify.

Firstly, we consider the distributions related to the loss variable $Y$. 
The likelihood function $p(\bY|\bPhi,\bZ)$ is given as follows:
$$
	p(\bY|\bPhi,\bZ)=\prod_{i=1}^n \prod_{k=1}^K (p_k(Y_i; \phi_k))^{I_k(Z_i)}=\prod_{k=1}^K \prod_{i:Z_i=k} p_k(Y_i; \phi_k),
$$
where $I_k(Z_i)=1$ if $Z_i=k$ else $I_k(Z_i)=0$.
The full conditional distribution of $\bPhi$ is given as follows:
$$p(\bPhi|\bY,\bD,\bZ,\bPsi,\btheta)=p(\bPhi|\bY,\bZ)\propto p(\bY|\bPhi,\bZ)p(\bPhi)=\prod_{k=1}^K q_k(\bphi_k;\bbeta_k) \prod_{i=1}^n \prod_{k=1}^K (p_k(Y_i; \phi_k))^{I_k(Z_i)}.$$
Therefore, the full conditional distribution of $\bphi_k$ is given as follows:
\begin{equation}\label{phi-full-conditional}
p(\bphi_k|\bY,\bZ)\propto q_k(\bphi_k;\bbeta_k) \prod_{i:Z_i=k} p_k(Y_i; \bphi_k).
\end{equation}
Note that when possible, we assume the conjugate prior $q_k$ for $\bphi_k$. For example, when the log-normal loss distribution is used, the conjugate priors for the mean and the variance are  the normal and inverse-gamma distribution, respectively.

Secondly, we consider the distributions related to the claim description text $\bD$.
The likelihood function of $\bD$ is given by
$$p(\bD|\bZ,\boldsymbol\Psi)=\prod_{i=1}^n \prod_{j=1}^{|\bD_i|}Dis(D_{i,j};\boldsymbol\psi_{Z_i})
=\prod_{i=1}^n \prod_{j=1}^{|\bD_i|}\prod_{v=1}^{|V|}(\psi_{Z_i, v})^{I_{w_v}(D_{i,j})}=\prod_{k=1}^K\prod_{v=1}^{|V|}(\psi_{k, v})^{\sum_{i:Z_i=k}\sum_{j=1}^{|\bD_i|}I_{w_v}(D_{i,j})},
$$
where $I_{w_v}(D_{i,j})=1$ if $D_{i,j}=w_v$ else $I_{w_v}(D_{i,j})=0$, and $\psi_{k,v}$ is the probability of occurring word $w_v$ in the topic $k$. 
We denote the number of occurrences of word $w_v$ in the $i$th document by a scalar  $N_{i,v}=N_{v}(\bD_i)=\sum_{j=1}^{|\bD_i|}I_{w_v}(D_{i,j})$ 
and all the occurrences in the $i$th document by a $|V|$-dimensional vector $\bN_i=\bN(\bD_i)=(N_{i,1},\ldots,N_{i,|V|})$.
It is clear that the sufficient statistics for $\boldsymbol\psi_k$ is the number of  occurrences of each word in topic $k$, $\sum_{i:Z_i=k} \bN_i$. 
The Dirichlet prior of $\bpsi_k$ is given by 
$$p(\bpsi_k)=Dir(\bpsi_k;\bgamma)=\frac{1}{C(\bgamma)}\prod_{v=1}^{|V|} (\psi_{k,v}) ^{\gamma_v-1},$$
where 
$$C(\bgamma)=\int_{\Delta} \prod_{v=1}^{|V|} \psi_{k,v}^{\gamma_v-1} d\bpsi_k=\frac{\prod_{v=1}^{|V|} \Gamma(\gamma_v)}{\Gamma(\sum_{v=1}^{|V|}\gamma_v)}.$$
Here $\Delta=\{(\psi_{k,1},\ldots,\psi_{k,|V|}):\sum_{v=1}^{|V|} \psi_{k,v}=1\}$ and $\Gamma$ is a generalization of the factorial function.
The full conditional distribution of $\bpsi_k$ is given as follows:
\begin{equation}\label{psi-full-conditional}
\begin{aligned}
	p(\boldsymbol \psi_k|\bD,\bZ)&\propto p(\bD|\bZ,\boldsymbol \Psi)p(\boldsymbol \Psi) \\
	&\propto \prod_{v=1}^{|V|}(\psi_{k, v})^{\sum_{i:Z_i=k}\sum_{j=1}^{|\bD_i|}I_{w_v}(D_{i,j})}\prod_{v=1}^{|V|}(\psi_{k, v})^{\gamma_v-1} \\
	&=\prod_{v=1}^{|V|}(\psi_{k, v})^{\gamma_v-1+\sum_{i:Z_i=k}N_{i,v}}.
\end{aligned}
\end{equation}
That is, $(\boldsymbol \psi_k|\bD,\bZ)$ follows a Dirichlet distribution with parameter $\boldsymbol\gamma + \sum_{i:Z_i=k} \bN_i$.
So  the hyper-parameters $\bgamma$ can be regarded as the pseudo-data.
The maximum a posteriori (MAP) estimate of $\psi_{k,v}$ is given by
$$\hat{\psi}_{k,v}=\frac{\gamma_v-1+\sum_{i:Z_i=k}N_{i,v}}{\sum_{v'=1}^{|V|} (\gamma_{v'}-1)+\sum_{v'=1}^{|V|}\sum_{i:Z_i=k}N_{i,v'}}.$$
If $\bgamma=(1,\ldots,1)$, then the MAP estimate is also the maximum likelihood estimate (MLE) for $\bpsi_k$.
In this paper, we set $\bgamma=(2,\ldots,2)$ which is referred to as Laplace smoothing. If a  word $w_v$ never occur in the topic $k$, the MAP estimate of $\psi_{k,v}$ is $1/|V|$.

Thirdly, we consider the distributions related to the latent variable $Z$.
The likelihood function of $\bZ$ is given by
$$p(\bZ|\btheta)=\prod_{i=1}^n Dis(Z_i;\btheta)=\prod_{i=1}^n \prod_{k=1}^K (\theta_{k})^{I_k(Z_i)}=\prod_{k=1}^K (\theta_k)^{\sum_{i=1}^n I_k(Z_i)},$$
where $I_k(Z_i)=1$ if $Z_i=k$ else $I_k(Z_i)=0$.
Therefore, the sufficient statistics for $\btheta$ is the number of samples belonging to each topic $\sum_{i=1}^n I_k(Z_i)$. We denote the number of samples with $Z_i=k$ by a scalar
$M_k=M_k(\bZ)=\sum_{i=1}^n I_k(Z_i)$ and all the numbers of samples in each topic by a $K$-dimension vector $\bM=\bM(\bZ)=(M_1,\ldots,M_K)$.
Similar to $\bpsi_k$, the Dirichlet prior is the conjugate prior for the discrete likelihood function and the full conditional distribution of $\btheta$ is given by:
\begin{equation}\label{theta-full-conditional}
\begin{aligned}
p(\btheta|\bZ)&\propto p(\bZ|\btheta)p(\btheta) \\
& \propto \left(\prod_{k=1}^K (\theta_k)^{M_k} \right) \left(\prod_{k=1}^K (\theta_k) ^{\alpha_k-1}\right) \\
& = \prod_{k=1}^K (\theta_k)^{M_k+\alpha_k-1}.
\end{aligned}
\end{equation}
That is, $(\btheta|\bZ)$ follows a Dirichlet distribution with parameter $\bM+\balpha$.
The hyper-parameters $\balpha$ can be regarded as the pseudo-data.
The maximum a posteriori (MAP) estimate of $\btheta$ is given by
$$\hat{\theta}_k=\frac{M_k+\alpha_k-1}{n+\sum_{k'=1}^K (\alpha_{k'}-1)}.$$
If $\balpha=(1,\ldots,1)$, then the MAP estimate is also the maximum likelihood estimate (MLE) for $\btheta$.
The full conditional distribution of $\bZ_i$ is given by
$$
\begin{aligned}
p(Z_i | \bY,\bD, \bZ_{-i},\bPhi,\btheta,\bPsi)& \propto p(Z_i | Y_i,\bD_i,\bPhi,\btheta,\bPsi) \\
&\propto 	p(\bPhi,\btheta,\bPsi,Z_i|Y_i,\bD_i) \\
& \propto p(Y_i|\bPhi,Z_i)p(\bD_i|\bPsi,Z_i)p(Z_i|\btheta)\\
&= \prod_{k=1}^K  (p_k(Y_i; \phi_k))^{I_k(Z_i)} \prod_{k=1}^K \left( \prod_{v=1}^{|V|}(\psi_{k, v})^{N_{i,v}}\right)^{I_k(Z_i)}  \prod_{k=1}^K (\theta_k)^{^{I_k(Z_i)}}
\end{aligned}
$$
Therefore, the full conditional distribution of $Z_i\in\{1,\ldots,K\}$ is given by 
\begin{equation*}
\Pr(Z_i=k|Y_i,\bD_i,\bphi_k,\theta_k,\bpsi_k)\propto \theta_kp_k(Y_i;\phi_k)\prod_{v=1}^{|V|}(\psi_{k,v})^{N_{i,v}},
\end{equation*}
which implies 
\begin{equation}\label{z-full-conditional}
\Pr(Z_i=k|Y_i,\bD_i,\bphi_k,\theta_k,\bpsi_k)=\frac{\theta_kp_k(Y_i;\phi_k)\prod_{v=1}^{|V|}(\psi_{k,v})^{N_{i,v}}}{\sum_{k'=1}^K\theta_{k'}p_{k'}(Y_i;\phi_{k'})\prod_{v=1}^{|V|}(\psi_{k',v})^{N_{i,v}}}.
\end{equation}

{Fourthly, the posterior predictive distribution of the claim loss $Y_{n+1}$ given the claim description $\bD_{n+1}$ is derived as follows:
	\begin{equation}\label{eq:pred1}
	\begin{aligned}
		&p(Y_{n+1}| \bD_{n+1}, \bY,\bD)\\
		=&
		\sum_{k=1}^K \int_{\btheta,\bPhi,\bPsi}p(Y_{n+1},Z_{n+1}=k,\btheta,\bPhi,\bPsi|\bD_{n+1}, \bY,\bD) d\btheta d\bPhi d\bPsi  \\
		=& \sum_{k=1}^K \int_{\btheta,\bPhi,\bPsi} p(Y_{n+1}|Z_{n+1}=k, \bphi_k) \Pr(Z_{n+1}=k| \bD_{n+1}, \btheta,\bPhi,\bPsi) p(\btheta,\bPhi,\bPsi|\bY,\bD)d\btheta d\bPhi d\bPsi  \\
		=&\sum_{k=1}^K \int_{\btheta,\bphi_k,\bpsi_k}  p_k(Y_{n+1}; \bphi_k)  \Pr(Z_{n+1}=k| \bD_{n+1}, \theta_k,\bpsi_k) p(\btheta,\bPhi,\bPsi|\bY,\bD)d\btheta d\bPhi d\bPsi.
	\end{aligned}
	\end{equation}
Here
$$\Pr(Z_{n+1}=k| \bD_{n+1}, \theta_k,\bpsi_k)\propto \theta_k\prod_{v=1}^{|V|}(\psi_{k,v})^{N_{n+1,v}},$$
which implies 
\begin{equation}\label{z-conditional-no-Y}
	\Pr(Z_i=k|\bD_i,\bphi_k,\theta_k,\bpsi_k)=\frac{\theta_k \prod_{v=1}^{|V|}(\psi_{k,v})^{N_{i,v}}}{\sum_{k'=1}^K\theta_{k'}\prod_{v=1}^{|V|}(\psi_{k',v})^{N_{i,v}}}.
\end{equation}
The integration and the summation can be  evaluated numerically in the Gibbs sampler which will be discussed later.

\subsection{The EM algorithm for the MAP estimate}

In this section we discuss the expectation-maximization (EM) algorithm for the maximum a posteriori (MAP) estimate of the parameters $\btheta,\bPhi,\bPsi$, which provides good initial values used in the Gibbs sampler.
The EM algorithm is an iterative method formally introduced by \citet{1977Maximum}, used for finding the maximum likelihood estimates (MLE) or the MAP estimates of parameters in the case of  missing data or latent variables. 
%The algorithm provides a systematic approach to deal with statistical model problems that involve latent variables or missing data, and it has been extensively studied and expanded upon. 
It is now applied to a variety of statistical models, including mixture models, hidden Markov models,   high-dimensional data analysis, complex dependency structures etc. 
One key advantage of the EM algorithm is its simplicity and elegance. It breaks down a complex optimization problem into two simpler steps: one involving the computation of an expectation, and the other involving optimization.

%Notice that \eqref{equ:lf_object1} contains a log of sums, which makes a maximization by partial derivatives computationally intractable. We consider a random vector of complete information $(S, \mathbf Z)$, where $S$ represents a random variable corresponding to the observed sample and $ Z_i=(Z_{i1},\ldots,Z_{iK}), Z_{ik}\in \{0,1\}, i = 1, ..., n$ is the latent component indicator variable, where $z_{ik}=1$ iff $z_i=k$ else $z_{ik}=0$. 
In the proposed model,  the likelihood function of the observed variable $(\bY,\bD)$  is complicated due to the latent variable $\bZ$.
The complete  posterior distribution of the parameters given the full information $(\bY, \bD, \bZ)$ is more tractable as follows:
\begin{equation*}
\begin{aligned}
&p(\bPhi,\btheta,\bPsi|\bY,\bD,\bZ)\\
\propto& p(\bY,\bD,\bZ|\bPhi,\btheta,\bPsi)p(\bPhi,\btheta,\bPsi) \\
\propto& p(\bY|\bPhi,\bZ)p(\bD|\bPsi,\bZ)p(\bZ|\btheta)p(\btheta)p(\bPhi)p(\bPsi)\\
\propto& \prod_{k=1}^K \left( q_k(\bphi_k;\bbeta_k) \prod_{i:Z_i=k} p_k(Y_i; \bphi_k)\right) \prod_{k=1}^K \left( \prod_{v=1}^{|V|}(\psi_{k, v})^{\gamma_v-1+\sum_{i:Z_i=k}N_{i,v}} \right) \prod_{k=1}^K (\theta_k)^{M_k+\alpha_k-1}\\
\propto& \prod_{k=1}^K \left( q_k(\bphi_k;\bbeta_k) \prod_{i:Z_i=k} p_k(Y_i; \bphi_k)  \prod_{v=1}^{|V|}(\psi_{k, v})^{\gamma_v-1+\sum_{i:Z_i=k}N_{i,v}} (\theta_k)^{M_k+\alpha_k-1} \right).
\end{aligned}
\end{equation*}
The complete logged posterior distribution is given by 
\begin{equation}\label{log-joint-posterior}
\begin{aligned}
&\log p(\bPhi,\btheta,\bPsi|\bY,\bD,\bZ)) \\
\propto  &\sum_{k=1}^K \left[\log q_k(\bphi_k;\bbeta_k)+\sum_{i:Z_i=k} \log p_k(Y_i;\bphi_k) + \sum_{v=1}^{|V|} \left(\gamma_v-1+\sum_{i:Z_i=k}N_{i,v}\right) \log \psi_{k,v}+ (M_k+\alpha_k-1)\log\theta_k\right],
\end{aligned}
\end{equation}
where the term $\sum_{i:Z_i=k} \log p_k(Y_i;\bphi_k), \sum_{i:Z_i=k}N_{i,v}$ and $M_k$ depends on the variable $\bZ$.
Equation \eqref{z-full-conditional} gives the full conditional distribution of $(Z_i=k|Y_i,\bD_i,\bphi_k,\theta_k,\bpsi_k)$, and
we denote the conditional probability of $Z_i=k$ given the parameters $\bPhi^{(t)},\btheta^{(t)},\bPsi^{(t)}$ by a scalar 
\begin{equation}\label{w=Ez}
w_{i,k}^{(t)}=\Pr(Z_i=k|Y_i,\bD_i,\theta_k^{(t)},\bphi_k^{(t)},\bpsi_k^{(t)}).
\end{equation}
We also denote a $n\times K$ matrix $\bW^{(t)}=(\bw_1^{(t)},\ldots,\bw_n^{(t)})$ where $\bw_i^{(t)}=(w_{i,1}^{(t)},\ldots,w_{i,K}^{(t)})$ is a $K$-dimensional vector.

The expectation of the complete logged likelihood \eqref{log-joint-posterior} with respect to the conditional distribution of $Z_i$ in \eqref{z-full-conditional} is as follows:
\begin{equation}\label{expectation-log-joint-posterior}
	\begin{aligned}
		&Q(\bPhi,\btheta,\bPsi|\bPhi^{(t)},\btheta^{(t)},\bPsi^{(t)})\\
		=& \mathbb{E}_{\bZ \sim p(Z|\bD,\bY,\btheta^{(t)},\bPhi^{(t)},\bPsi^{(t)})} \log p(\bPhi,\btheta,\bPsi|\bY,\bD,\bZ)) \\
		\propto & \sum_{k=1}^K \left( 
		\underbrace{\log q_k(\bphi_k;\bbeta_k)+\sum_{i:=1}^n w_{i,k}^{(t)}\log p_k(Y_i;\bphi_k)}_\text{(1)}+\underbrace{ \sum_{v=1}^{|V|} \left(\gamma_v-1+\sum_{i=1}^n w_{i,k}^{(t)}N_{i,v}\right) \log \psi_{k,v}}_\text{(2)}+ \right.\\ & \left. \underbrace{\left(\sum_{i=1}^n w_{i,k}^{(t)}+\alpha_k-1\right)\log\theta_k}_\text{(3)}\right),
	\end{aligned}
\end{equation}
where $(1)$ corresponds to the logged posterior distribution of  $(\bphi_k|\bY)$ with weights of $\bW^{(t)}$, $(2)$ corresponds to the logged posterior distribution of $(\bpsi_k|\bD)$ with weights of $\bW^{(t)}$, and $(3)$ corresponds to the logged posterior distribution of $(\theta|\bW^{(t)})$.
The parameters $\bPhi,\btheta,\bPsi$ maximizing function \eqref{expectation-log-joint-posterior} can be derived as:
\begin{align}
	\bphi_k^{(t+1)}&=\underset{\bphi_k}{\arg\max} \left[ \log q_k(\bphi_k;\bbeta_k)+\sum_{i:=1}^n w_{i,k}^{(t)}\log p_k(Y_i;\bphi_k)\right], ~ k=1,\ldots,K, \label{m-phi} \\ 
	\psi_{k,v}^{(t+1)}&= \frac{\gamma_v-1+\sum_{i=1}^n w_{i,k}^{(t)}N_{i,v}}{\sum_{v'=1}^{|V|} (\gamma_{v'}-1)+\sum_{v'=1}^{|V|}\sum_{i=1}^n w_{i,k}^{(t)}N_{i,v'}},~ k=1,\ldots,K, v=1,\ldots,|V|, \label{m-psi} \\
	\theta_{k}^{(t+1)}&=\frac{\sum_{i=1}^n w_{i,k}^{(t)}+\alpha_k-1}{n+\sum_{k'=1}^K (\alpha_{k'}-1)}. \label{m-theta}
\end{align}
Note that $\bphi_k^{(t+1)}$ is the MAP estimate when fitting $\bY$ with weights $\bW$ to the loss distribution $p_k(\bphi_k)$ with the prior $q_k$.
For example, the log-normal loss distribution $p_k(\bphi_k)=LN(\mu_k,\sigma_k)$ with the non-informative prior $q_k$ leads to the following MAP estimate (also the MLE): 
\begin{align}
	\hat\mu^{(t+1)}_k &= \frac{\sum_{i=1}^nw_{i,k}^{(t)}\log Y_i}{\sum_{i=1}^nw_{i,k}^{(t)}},\\ 
	\hat\sigma_k^{(t+1)} &= \sqrt{\frac{\sum_{i=1}^nw_{i,k}^{(t)}\left(\log Y_i-\hat\mu_k^{(t+1)}\right)^2}{\sum_{i=1}^nw_{i,k}^{(t)}}}
\end{align}
The Pareto distribution will be discussed in the next subsection. 
We also consider the general beta of the second kind (GB2) loss distribution\footnote{The general beta of the second kind (GB2) distribution nests a number of important distributions such as 
log-normal, Weibull, gamma, Lomax, Chi-square, half-normal, exponential, asymmetric log-Laplace, etc.}. 
%It favors, in particular, the modelling of loss reserve data which often exhibits heavy-tailed behaviour from long tail business classes. 
The GB2 distribution consists of four parameters and its probability density function is specified in Appendix \ref{appendix:loss}.
We will use R package GB2 to perform the maximization in equation \eqref{m-phi} if the GB2 loss distribution is used. 

The detail of the EM algorithm  is shown in Algorithm \ref{alg:em}.
The E-step and M-step alternate until the relative increase in the logged posterior distribution at two consecutive iterations is no bigger than a small pre-specified tolerance value.
\begin{algorithm}[h!]
	\caption{The EM algorithm for the MAP estimates.}
	\label{alg:em}
	\begin{algorithmic}[1]
		\STATE Input: Loss $\bY=(Y_1,\cdots,Y_n)$, claim description text $\bD = (\bD_1,\cdots,\bD_n)$ and the number of clusters $K$.
	\STATE	Initialization. 
For each sample $i=1,\ldots,n$, simulate a $Z_i^{(-1)}$ from a discrete distribution with even probabilities. 
Given $\bW^{(-1)}=\bZ^{(-1)}$,  set $\bPhi^{(0)},\bPsi^{(0)},\btheta^{(0)}$ according to equations \eqref{m-phi}, \eqref{m-psi} and \eqref{m-theta}.
		\FOR{$t=1$ to $T$}
		\STATE E-step. Calculate $\bW^{(t-1)}$ by equations \eqref{z-full-conditional} and \eqref{w=Ez} with $\theta_k=\theta_k^{(t-1)},\bphi_k=\bphi_k^{(t-1)}, \bpsi_k=\bpsi_k^{(t-1)}$.
		\STATE M-step. Given $\bW^{(t-1)}$, calculate $\bPhi^{(t)},\bPsi^{(t)},\btheta^{(t)}$ by the equations \eqref{m-phi}, \eqref{m-psi} and \eqref{m-theta}.
	\ENDFOR
	\end{algorithmic}
\end{algorithm}

\subsection{The Gibbs sampler for the posterior distribution}

A full Bayesian posterior analysis always includes estimating the posterior distribution of parameters which qualifies the parameter estimation uncertainty.
The Metropolis-Hastings (MH) algorithm is a general Monte Carlo Markov chain (MCMC) method, originally proposed by \citet{Metropolis-1953} to simulate from the posterior distribution of parameters. 
Then the algorithm was extended by \citet{hastings1970monte}, which is  generalized to allow for different types of proposal distributions. 
Gibbs sampler is a special version of the MH algorithm, first introduced by \citet{geman1984stochastic}.
An advantage of Gibbs sampler is that it does not require an accept-reject criterion, as the full conditional distribution is chosen as the proposal distribution and the acceptance rate is always 1. 

In this section, we introduce the  MH-within-Gibbs sampler algorithm for the LDMM model. 
The MH-within-Gibbs  sampler combines the advantages of the two aforementioned methods. 
On one hand, if the  full conditional distributions of some variables are  recognizable, then the Gibbs sampling steps are used; 
on the other hand, if the full conditional distributions of some variables are not recognizable, then the MH steps are used.

In the proposed LDMM model, since the conjugate priors for  $\bPsi$ and $\btheta$ are used, the full conditional distributions \eqref{psi-full-conditional} and \eqref{theta-full-conditional} are  recognizable which facilitates the Gibbs sampler.
While for $\bPhi$, when the conjugate priors are not available, the MH algorithm is implemented to simulate from the distribution \eqref{phi-full-conditional}.    
Note that we also simulate the latent variable $Z_i$ from its full conditional distribution \eqref{z-full-conditional}.
The details of the MH-within-Gibbs sampler algorithm is shown in Algorithm \ref{alg:gs}.

\begin{algorithm}[h!]
	\caption{The MH-within-Gibbs sampler  algorithm for the posterior distribution.}
	\label{alg:gs}
	\begin{algorithmic}[1]
		\STATE Input: Loss $\bY=(Y_1,\cdots,Y_n)$, claim description text $\bD = (\bD_1,\cdots,\bD_n)$, the number of clusters $K$, the MAP estimates ${\bPhi}^{(T)}, {\bPsi}^{(T)}, {\btheta}^{(T)}$ from  and the last updated weights $\bW^{(T-1)}$ from Algorithm \ref{alg:em}.
\STATE	Initialization. 
Simulate a $Z_i^{[0]}$ from a discrete distribution with probabilities $\bw_i^{(T-1)}$ for each sample $i=1,\ldots,n$. 
Set ${\bPhi}^{[0]}={\bPhi}^{(T)}, {\bPsi}^{[0]}={\bPsi}^{(T)}, \btheta^{[0]}={\btheta}^{(T)}$.
\FOR{$t=1$ to $T$}
\STATE Simulate $\bphi_k^{[t]}$ from the full conditional distribution \eqref{phi-full-conditional} $p(\bphi_i|\bY,\bZ^{[t-1]})$ for $k=1,\ldots,K$. Note that if the full conditional distribution \eqref{phi-full-conditional} is not recognizable then a MH step is used.
\STATE Simulate $\bpsi_k^{[t]}$ from the Dirichlet distribution   \eqref{psi-full-conditional} with parameter $\bgamma+\sum_{i:Z^{[t-1]}_i=k}\bN_i$ for $k=1,\ldots,K$.
\STATE Simulate $\btheta^{[t]}$ from the Dirichlet distribution \eqref{theta-full-conditional} with parameter $\bM^{[t-1]}+\balpha$, where $\bM^{[t-1]}=(M_1^{[t-1]},\ldots, M_K^{[t-1]})=(\sum_{i=1}^n I_1(Z_i^{[t-1]}),\ldots,\sum_{i=1}^n I_K(Z_i^{[t-1]}))$.
\STATE Simulate $\bZ^{[t]}\in\{1,\ldots,K\}$ from the discrete distribution \eqref{z-full-conditional} $p(Z_i^{[t]}=k|Y_i,\bD_i,\bphi_k^{[t-1]},\theta_k^{[t-1]},\bpsi_k^{[t-1]})$
\ENDFOR
	\end{algorithmic}
\end{algorithm}

{We list the conjugate priors for the log-normal and Pareto loss distribution as follows.} 
\begin{itemize}
	\item Log-normal distribution $Y_i\sim LN(\mu, \sigma^2)$: the conjugate prior and posterior distribution are normal-inverse gamma, given by \eqref{equ:conjugate_ln1} and \eqref{equ:conjugate_ln2}:
	\begin{equation}\label{equ:conjugate_ln1}
		\mu, \sigma^2\sim N\mbox{-}\Gamma^{-1}(\mu_0, r , a , b ),
	\end{equation}
	\begin{equation}\label{equ:conjugate_ln2}
		\mu, \sigma^2|\bY \sim N\mbox{-}\Gamma^{-1}\left(\frac{r \mu_0+n\overline{\log \bY}}{r +n}, \frac{r+n}{\sigma^2}, a +\frac{n}{2}, b +\frac{1}{2}\sum_{i=1}^n(\log Y_i-\overline{\log \bY})^2+\frac{nr }{r +n}\frac{(\overline{\log \bY}-\mu_0)^2}{2}\right),
	\end{equation}
	where  $\overline{\log \bY}$ denotes the sample mean of  $(\log Y_i)_{i=1}^n$. The density function of the normal-inverse gamma distribution $N\mbox{-}\Gamma^{-1}$ is specified in Appendix \ref{appendix:loss}.
	
	\item Pareto distribution with known minimum $\sigma$: the conjugate prior and posterior distribution is gamma distribution, given by \eqref{equ:conjugate_pareto1} and \eqref{equ:conjugate_pareto2}:
	\begin{equation}\label{equ:conjugate_pareto1}
		r\sim \Gamma(a, b),
	\end{equation}
	\begin{equation}\label{equ:conjugate_pareto2}
		r|\bY \sim \Gamma\left(a+n, b+\sum_{i=1}^n\ln \frac{Y_i}{\sigma}\right).
	\end{equation}
\end{itemize}

\subsection{Model selection}\label{sec:model-selection}

In the LDMM model, we need to determine the number of components and the component loss distributions. 
Therefore, model selection is crucial for the LDMM model.
In this section, we discuss four metrics: the deviance information criterion (DIC), Wasserstein distance, perplexity and stability.
{The first one is used to evaluate the goodness-of-fit of the entire model and the second one is used to evaluate the goodness-of-fit of the finite mixture model for the claim loss, while the last two are used to evaluate the performance of the DMM for the claim description (see Figure \ref{fig:LDMM}).}

The DIC, proposed by \citet{2002Bayesian}, is a hierarchical modeling generalization of the Akaike Information Criterion (AIC), which is an asymptotic approximation to AIC as the sample size becomes large. 
The DIC is particularly useful in Bayesian model selection  where the posterior distribution of  parameters have been sampled by the Markov chain Monte Carlo (MCMC) methods.
The DIC is defined as 
$$\mbox{DIC}=p_D+\overline{D(\theta)}
\text{~or equivalently as~}
\mbox{DIC}=D(\bar\theta)+2p_D,$$
where a slight abuse of $\theta$  indicates all the parameters, $p_D = \overline{D(\theta)}-D(\bar\theta)$ is the  effective number of  parameters,  $\bar\theta$ is the posterior expectation of $\theta$, $\overline{D(\theta)}$ is the posterior expectation of $D(\theta)$, 
and 
$$
\begin{aligned}
D(\theta)=&-2\log p(\bY,\bD| \bPhi,\bPsi,\btheta)\\
=& -2 \sum_{i=1}^n \log p(Y_i,\bD_i|\bPhi,\bPsi,\btheta)\\
=& -2\sum_{i=1}^n  \log\left( \sum_{k=1}^K p(Y_i,\bD_i,Z_i=k|\bPhi,\bPsi,\btheta)\right) \\
=& -2\sum_{i=1}^n  \log\left( \sum_{k=1}^K p(Y_i,\bD_i|Z_i=k,\bphi_k,\bpsi_k)\Pr(Z_i=k|\btheta)\right)\\
=& -2\sum_{i=1}^n  \log\left( \sum_{k=1}^K \left(\theta_k p_k(Y_i;\bphi_k)\prod_{v=1}^{|V|}\psi_{k,v}^{N_{i,v}}\right) \right).
\end{aligned}
$$
Note that the proposed model contains the latent variable $Z$, so the above deviance are called  observed-data deviance or the integrated deviance.
The larger the effective number of parameters is, the easier overfitting, and  the more penalty in the DIC. 
Models with smaller DIC are preferred. 

The Wasserstein distance (also called as Earth Mover’s distance), proposed by \citet{2000The} arising from the idea of optimal transport, 
is being used more and more frequently in statistics and machine learning. 
It is a distance metric between two probability distributions on a given metric space. 
Intuitively, it can be seen as the minimum work needed to transform one distribution to another, where work is defined as the product of mass of the distribution that has to be moved and the distance to be moved. 
Wasserstein distance with index $\rho\ge 1$ between two probability measure $\xi$ and $\nu$ is defined by
\begin{align}
	\mathcal W_\rho(\xi,\nu) = \inf_{\pi\in\Pi(\xi,\nu)}\left(\int_{\mathbb R\times \mathbb R}|x-y|^{\rho}\pi(dx,dy)\right)^{1/\rho},
\end{align}
where $\xi$ and $\nu$ are the probability measures on the real line, $\Pi(\xi,\nu)$ is the set of couplings between $\xi$ and $\nu$, that is $\pi\in\Pi(\xi,\nu)$ iff $\pi$ is a measure on $\mathbb R\times \mathbb R$ with first marginal $\xi$ and second marginal $\nu$. 
When dimension $d=1$, the distance has a closed form:
\begin{align}
	\label{align:wassertein}
	\mathcal{W}_\rho(\xi,\nu) = \left(\int_{(0,1)}\left|F_\xi^{-1}(u)-F_\nu^{-1}(u)\right|^\rho du\right)^{1/\rho},
\end{align}
where $F_\eta^{-1}(u) = \inf\{x\in\mathbb R: \eta((-\infty,x])\ge u\}$ denotes the quantile function of a probability measure $\eta$ on $\mathbb R$. 
Further, if $\xi$ is the empirical distribution of a dataset $u_1,...u_n$ and $\nu$ is the empirical distribution of another dataset $v_1,...,v_n$ of the same size, then the distance takes a very simple function of the order statistics:
\begin{align}
	\mathcal{W}_\rho(\xi,\nu) = \left(\frac{1}{n}\sum_{i=1}^n\|u_{(i)}-v_{(i)}\|^\rho\right)^{1/\rho}.
\end{align}
In this paper, we use the Wasserstein distance with index $\rho = 1$ to evaluate the difference between the estimated mixture of loss distributions  and the empirical loss distribution. 

Perplexity, proposed by \citet{2014Adaptive}, is a metric to evaluate the performance of topic model. 
It measures how well the model predicts unseen or held-out documents. 
Lower perplexity scores indicate that the model can better predict the words in unseen documents, suggesting a better understanding of the underlying topics.
The perplexity is calculated as
\begin{align}
\label{align:perplexity}
    \mbox{perplexity} = \exp\left\{-\frac{\sum_{i\in\mathcal{I}_{test}}\sum_{v=1}^{|V|}\log\left(\sum_{k = 1}^K(\psi_{k, v})^{N_{i,v}}\Pr(Z_i=k)\right)}{\sum_{i\in\mathcal{I}_{test}} |\bD_i|}\right\},
\end{align}
where $\mathcal{I}_{test}$ denotes the set of test indices,  $|\bD_i|$ is the number of words in the $i$th document and $\Pr(Z_i=k)$ is  given by \eqref{z-full-conditional}.
%\begin{equation}
%\begin{aligned}
%\label{align:perplexity1}
%\Pr(Z_i=k|\bD_i,\btheta,\bPsi)=&
%\frac{p(Z_i=k,\bD_i|\btheta,\bPsi)}{p(\bD_i|\btheta,\bPsi)}\\
%=&
%\frac{p(\bD_i|Z_i=k,\bPsi)\Pr(Z_i=k|\btheta)}{\sum_{k'=1}^K p(\bD_i|Z_i=k',\bPsi)\Pr(Z_i=k'|\btheta)} \\
%=&\frac{\theta_k\prod_{v=1}^{|V|} \psi_{k,v}^{N_{i,v}}}{\sum_{k'=1}^K \theta_{k'}\prod_{v=1}^{|V|} \psi_{k',v}^{N_{i,v}}}
%\end{aligned}
%\end{equation}

\citet{XingLinzi2018APVi} defines a metric called topic stability which measures the degree to which a topic’s parameters change during sampling. 
The parameters related to a topic $k$ are the $|V|$-dimensional word distribution vector $\bpsi_k$. 
The stability of parameters for topic $k$ is defined as:
\begin{align}
    \mbox{stability}_k= \frac{1}{T}\sum_{t=1}^T sim(\bpsi_k^{[t]},\bar\bpsi_k),
\end{align}
where \textit{sim} is a vector similarity function and $\bar\bphi_k$ is the posterior mean of $\bphi_k$. This paper experiments with two similarity metrics: Euclidean distance and KL-divergence (see Table \ref{tab:lwwptop20_stablility} in Appendix \ref{appendix:stability}).

\subsection{Posterior predictive distribution and risk measures}

In the risk management of  reported but not settled (RBNS) claims, we need to estimate their posterior predictive distribution and risk measures given the claims description. 
We discuss two commonly used risk measures, value-at-risk (VaR) and conditional tail expectation (CTE), also known as expected shortfall (ES). 
In actuarial practice, insurance companies may use both VaR and CTE to evaluate potential claims losses and set up the prudential capital.

The posterior predictive distribution of the claim loss $Y_{n+1}$ given the claim description $\bD_{n+1}$ is derived in equation \eqref{eq:pred1}. 
The integration and the summation in equation \eqref{eq:pred1} can be  evaluated numerically using the sampled parameters in Algorithm \ref{alg:gs}. The simulation procedure is detailed in Algorithm \ref{alg:risk_sampling}.

\begin{algorithm}[h!]
	\small\setstretch{1}
	\caption{\textbf{The posterior predictive distribution of the RBNS claims}}
	\label{alg:risk_sampling}
	\begin{algorithmic}[1]
		\STATE Input: Simulated parameters $\btheta^{[1:T]},\bPhi^{[1:T]},\bPsi^{[1:T]}$ from
		Algorithm \ref{alg:gs}.
		\FOR{$t=1$ to $T$}
		\STATE Simulate $Z_{n+1}^{[t]}\in\{1,\ldots,K\}$ from the discrete distribution \eqref{z-conditional-no-Y}.
		\STATE Simulate $Y_{n+1}^{[t]}$ from the loss distribution $p_k(Y_{n+1};\bphi_k)$, where $k=Z_{n+1}^{[t]}$.
		\ENDFOR
	\end{algorithmic}
\end{algorithm}

With the simulated samples $Y_{n+1}^{[1:T]}$, we can estimate the VaR and CTE directly.
VaR  quantifies the maximum loss that a portfolio might suffer over a given time frame, at a given confidence level (for example, 95\% or 99\%):
\begin{equation}\label{var}
    VaR_{n+1}(\alpha) = \inf \left\{y \in Y_{n+1}^{[1:T]}: \Pr(L > y) \leq 1 - \alpha\right\},
\end{equation}
where  $L$ follows the empirical distribution of $Y_{n+1}^{[1:T]}$.
CTE is a risk measure that considers the average of losses above the VaR threshold. 
Simply put, CTE is the conditional average of losses exceeding the VaR threshold:
\begin{equation}\label{cte}
    CTE_{n+1}(\alpha)  =\frac{\sum_{t:Y_{n+1}^{[t]}>VaR_{n+1}(\alpha) } Y_{n+1}^{[t]}}{\left|\left\{y \in Y_{n+1}^{[1:T]}: y>VaR_{n+1}(\alpha) \right\}\right|},
\end{equation}
where the denominator is the number of simulated samples larger than $VaR_{n+1}(\alpha)$.

\section{Experimental study}
In this section, we illustrate the proposed methods by  analyzing  a highly realistic synthetic data.
This dataset includes 90,000 realistic, synthetically generated worker compensation insurance policies, all of which have had an accident \citep{actuarial-loss-estimation}. 

\subsection{Data description}

 Table \ref{tab:rawdata} lists  the first four records in the dataset. 
 We have two variables, the claim amount and the textual claim description.
 Intuitively, the claim amount should be closely related to the claim description.
 For example, a claim of injuries to knee tend to be more severe than that of  a injured finger.
 \begin{table*}[h!]
 	\renewcommand\arraystretch{1.1}
 	\centering
 	\caption{The first four records in the dataset.}
 	\begin{tabular}{ccc}
 		\hline
 		claim amount && textual claim description\\
 		\hline
 		4748.20 && lifting type injury to right arm and wrist injury\\
 		6326.27 && stepped around crates and truck tray fracture left forearm\\
 		2293.94 && cut on sharp edge cut left thumb\\
 		1786.49 && digging lower back lower back strain\\
 		\hline
 	\end{tabular}
 	\label{tab:rawdata}
 \end{table*}

Table \ref{tab:rawdata_loss} and Fig \ref{fig:rawdata_loss} present descriptive statistics and the histogram of the claims amount. 
The five most extreme losses in the right tail are indicated by arrows.
The distribution of claims amount is characterized by clear multi-modality, right skews and thick tails. 
{We split the dataset into a training dataset (80\%) and a test dataset (20\%) in a stratified fashion w.r.t. the claim amount, i.e., the distributions of claim amount are similar in the training and test datasets.}

\begin{table*}[h!]
    \renewcommand\arraystretch{1.1}
    \centering
    \caption{The descriptive statistics of the claims amount.}
    \begin{tabular}{ccccccc}
    \hline
    Mean & Median & St. Dev & Skewness & Kurtosis & Minimum & Maximum\\
    \hline
    $11,003.37$ & $11,003.37$ & $33,390.99$ & 37.55 & $3,940.35$ & 121.89 & $4,027,135.00$\\
    \hline
    \end{tabular}
    \label{tab:rawdata_loss}
\end{table*}

\begin{figure}[h!]
	\centering
    \begin{minipage}{1\linewidth}
		\centering
		\includegraphics[width=0.8\linewidth]{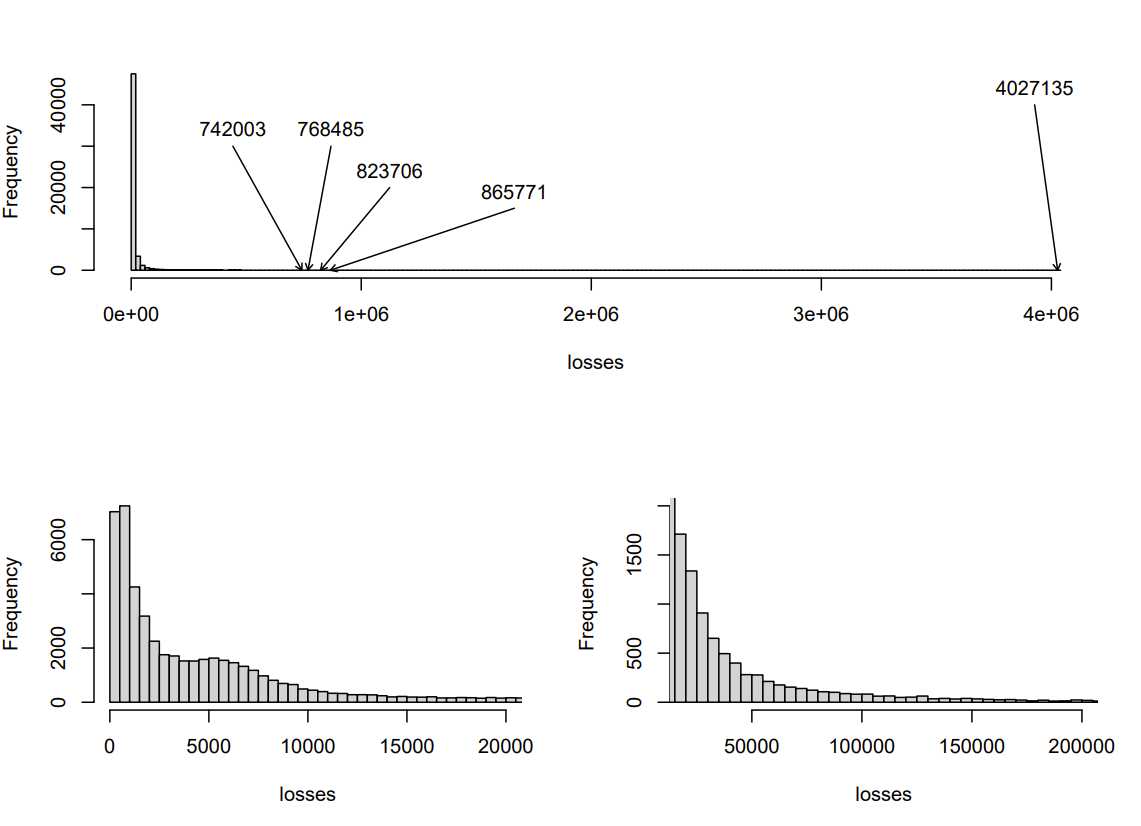}
		\caption{The distribution of the claims amount.}
		\label{fig:rawdata_loss}%文中引用该图片代号
	\end{minipage}
\end{figure}

Fig \ref{fig:rawdata_wordcount} shows that the claim description is a typical short text,  the number of words ranging from 1 to 14 with 6-8 words accounting for 77\% of the total. 
We implement the routine of pre-processing the textual data, e.g. changing to lower case, stemming and lemmatization, removing the stop words etc.
Based on the occurrence frequency of each word (called term frequency, TF), 
Figure \ref{fig:word-clouds} shows the word cloud which indicates that  the words \textit{right} and \textit{lower} have the highest TFs.
However, those words with high TFs  do not necessarily have the significance of distinguishing claims, e.g. right, left.
A commonly-used improvement is to set the inverse document frequency (IDF) as the weights when counting the occurrence of each word, named as TF-IDF.
It is well known that the TF-IDF can distinguish clams better than the TF.
Figure \ref{fig:word-clouds} shows the word cloud based on the TF-IDF. 
The words with the highest TF-IDF include \textit{back}, \textit{strain}, \textit{finger}, \textit{placement}, \textit{lifting}, \textit{shoulder}, etc.

\begin{figure}[h!]
	\centering
    \begin{minipage}{1\linewidth}
		\centering
		\includegraphics[width=0.8\linewidth]{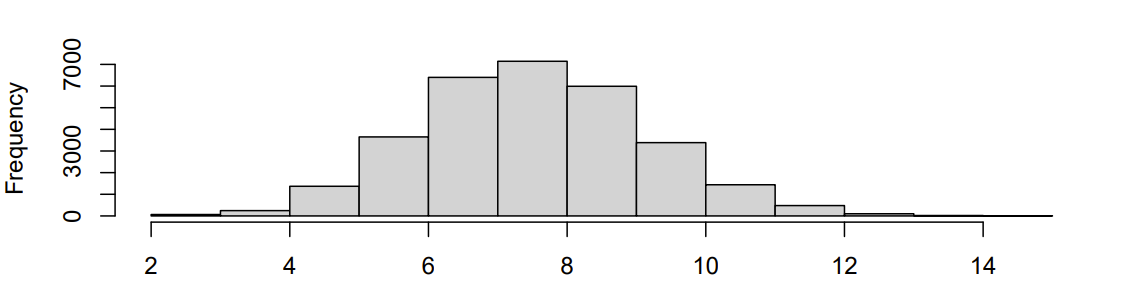}
		\caption{Histogram of word counts of the claim description.}
		\label{fig:rawdata_wordcount}
	\end{minipage}
\end{figure}

\begin{figure}[h!]
	\centering
    \begin{minipage}{0.45\linewidth}
		\centering
		\includegraphics[width=0.95\linewidth]{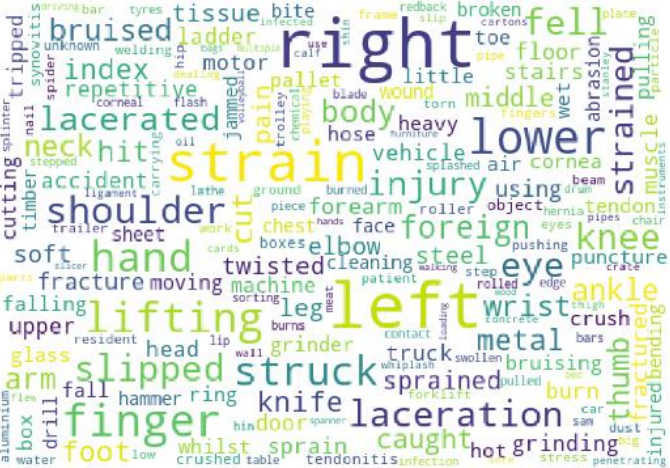}
	\end{minipage}
\begin{minipage}{0.45\linewidth}
	\centering
	\includegraphics[width=0.95\linewidth]{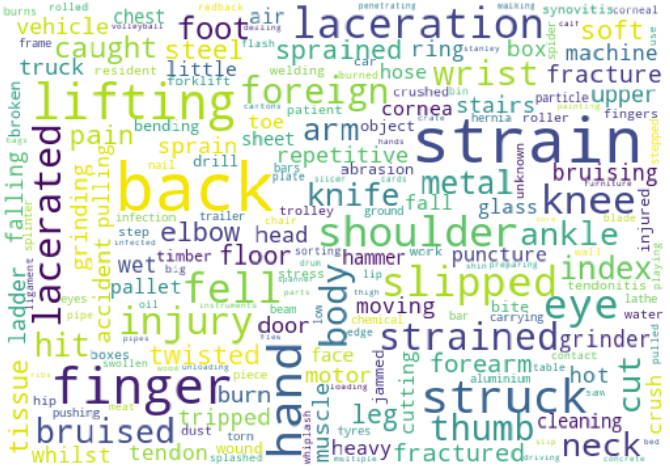}
\end{minipage}
\caption{Word clouds using the TF (left)  and TF-IDF (right).}\label{fig:word-clouds}
\end{figure}

\subsection{Parameter estimation}

In the following LDMM models, we use four loss distributions,
log-normal (LN),  GB2  and Pareto (P) distributions.
In the following, we consider mixtures of the same loss distribution family including Models 2LN,  $\ldots$, 6LN, 2GB2, $\ldots$, 6GB2, 2P, 3P, 4P.
 We also consider mixtures of different loss distribution families including Models 1LN1P, 2LN1P, 1LN2P, 3LN1P, 2LN2P, 1LN3P.
The MAP estimates and the posterior distribution of parameters can be obtained by Algorithms \ref{alg:em} and \ref{alg:gs}, respectively.
Note that how to determine the optimal number of components and select the component loss distributions are still open questions.
Here we pre-specify the number of components and component loss distributions, then compare the alternative models in terms of four metrics discussed in Section \ref{sec:model-selection}.

\subsubsection{MAP estimates}

We specify the number of components as $K=2,3,4,5,6$ and select the component loss distribution as either log-normal or GB2, resulting in 10 LDMM models (denoted by Models 2LN,  $\ldots$, 6LN, 2GB2, $\ldots$, 6GB2).
Figure \ref{fig:em_converge} shows the convergence  of Algorithm \ref{alg:em} for Models 2LN and 2GB2, respectively. 
When the number of EM iterations approaches to 50, we obtain the MAP estimates, which are listed in Tables \ref{tab:emLDMM_GB2_parameter} for Models 2GB2, $\ldots$, 6GB2. 
The parameters of loss distribution vary greatly in different components, reflecting different claim characteristics in different components. 

{We list the top-5 words with the highest occurrence probabilities $\bpsi_k$  in each topic $k$ and find that those words explain the characteristics of each component loss distribution to some extent. 
	For example, Model 2GB2 distinguishes two types of claim: the first severe ones include injuries to back and shoulder and the second mild ones include injuries to finger, hand and eye.}
%\blue{Top-5 words are not good measure since some words appear in several groups.}

\begin{figure}[h!]
	\centering
	\includegraphics[width=0.48\linewidth]{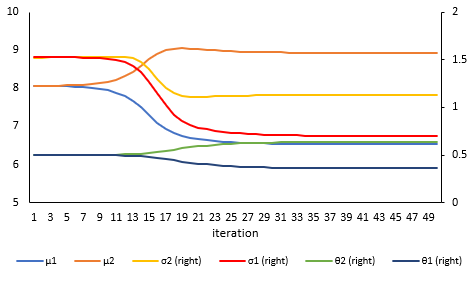}
		\includegraphics[width=0.48\linewidth]{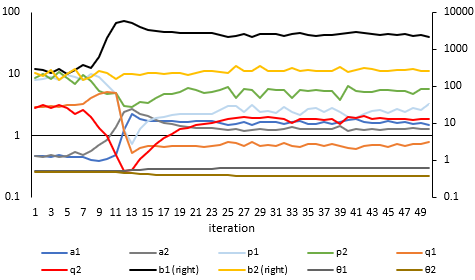}	
	\caption{The convergence of Algorithm \ref{alg:em} for Models 2LN and 2GB2.}	\label{fig:em_converge}%文中引用该图片代号
\end{figure}

\begin{table*}[h!]
    \small
    \renewcommand\arraystretch{1.2}
    \centering
    \caption{The MAP estimates of parameters and the top-5 words in the LDMM models with GB2 loss distributions.}
   \begin{tabular}{ccccccc}
   	\hline\hline
   	Models                  & $\theta$ & $a$   & $b$   & $p$   & $q$   & Top-5 words                               \\ \hline
   	\multirow{2}{*}{{2GB2}} & 0.612    & 1.565 & 2410  & 2.711 & 0.745 & back, strain, lifting, shoulder, fell     \\
   	& 0.3872   & 1.250 & 256   & 5.713 & 1.904 & finger, struck, hand, eye, laceration     \\ \hline
   	\multirow{3}{*}{{3GB2}} & 0.5492   & 5.386 & 4627  & 0.575 & 0.189 & back, strain, lifting, shoulder, fell     \\
   	& 0.1482   & 1.995 & 1544  & 3.924 & 3.389 & hand, finger, struck, slipped, fell       \\
   	& 0.3012   & 1.621 & 242   & 5.169 & 1.620 & finger, eye, struck, hand, foreign        \\ \hline
   	\multirow{4}{*}{{4GB2}} & 0.5432   & 6.094 & 4705  & 0.520 & 0.166 & back, strain, lifting, shoulder, fell     \\
   	& 0.0812   & 3.036 & 2432  & 2.367 & 3.342 & hand, finger, slipped, struck, fell       \\
   	& 0.0862   & 3.638 & 1046  & 2.593 & 1.429 & hand, finger, struck, laceration, fell    \\
   	& 0.2872   & 1.703 & 236   & 5.127 & 1.549 & finger, eye, struck, hand, foreign        \\ \hline
   	\multirow{5}{*}{{5GB2}} & 0.2952   & 2.229 & 322   & 2.596 & 1.043 & finger, eye, struck, hand, foreign        \\
   	& 0.1912   & 1.192 & 13880 & 3.020 & 1.639 & strain, back, lifting, shoulder, knee     \\
   	& 0.3522   & 4.130 & 6087  & 0.746 & 0.868 & back, strain, lifting, shoulder, strained \\
   	& 0.1002   & 3.217 & 1199  & 2.746 & 1.838 & hand, finger, struck, fell, laceration    \\
   	& 0.0612   & 4.091 & 2980  & 1.331 & 3.096 & hand, finger, slipped, struck, fell       \\ \hline
   	\multirow{6}{*}{{6GB2}} & 0.2762   & 1.953 & 267   & 3.850 & 1.312 & finger, eye, struck, hand, foreign        \\
   	& 0.0612   & 5.147 & 1584  & 2.190 & 1.549 & hand, finger, struck, slipped, fell       \\
   	& 0.0482   & 5.813 & 3325  & 0.908 & 2.798 & hand, finger, slipped, fell, struck       \\
   	& 0.3432   & 4.214 & 6055  & 0.758 & 0.873 & back, strain, lifting, shoulder, strained \\
   	& 0.0742   & 4.036 & 814   & 3.298 & 1.123 & hand, finger, struck, laceration, fell    \\
   	& 0.1952   & 1.224 & 14779 & 2.683 & 1.580 & strain, back, lifting, shoulder, knee     \\ \hline\hline
   \end{tabular}
    \label{tab:emLDMM_GB2_parameter}
\end{table*}

With the MAP estimates of parameters, 
Table \ref{tab:emLDMM_gb2_performance} compares those 10 LDMM models  in terms of negative log-likelihood (NLL), AIC, BIC, perplexity and Wassertein distance. 
Note that the claims amount always shows a high degree of positive skewness and thick tail, 
however, the Wassertein distance assumes the symmetry of the competing distributions.
Therefore, the claims amount and the fitted loss distributions are right-truncated the at 70\%, 80\%, and 90\% before calculating the Wassertein distance.
In this case, the Wassertein distance measures the goodness-of-fit in the central part of the loss distribution. 
The bolded numbers represent the best model under the corresponding metric. 
%It shows that in general  the LDMM models with GB2 loss distributions are better than those with log-normal loss distributions.
It seems that $K=3,4$ components are sufficient to capture the complexity in the claims amount distribution and when $K=6$ the model is overfitted with higher AIC, BIC and the test Wassertein distance.

\begin{table*}[h!]
	\small
	\renewcommand\arraystretch{1.1}
	\centering
	\caption{Comparison of 10 LDMM models at the MAP estimates.}
\begin{tabular}{cccccccccc}
	\hline
	\multirow{2}{*}{Models} & \multirow{2}{*}{NLL}            & \multirow{2}{*}{AIC}            & \multirow{2}{*}{AIC Rank}       & \multirow{2}{*}{BIC}            & \multirow{2}{*}{BIC Rank}       & \multirow{2}{*}{perplexity}      & \multicolumn{3}{c}{Wassertein distance} \\ \cline{8-10} 
	&                                 &                                 &                             &                                 &                             &                                  &           & Training         & Test          \\ \hline
	\multirow{3}{*}{2LN}  & \multirow{3}{*}{12.43}          & \multirow{3}{*}{25.20}          & \multirow{3}{*}{2}          & \multirow{3}{*}{\textbf{26.65}} & \multirow{3}{*}{\textbf{1}} & \multirow{3}{*}{1.5933}          & 70\%      & 163.8405              & 176.6555              \\
	&                                 &                                 &                             &                                 &                             &                                  & 80\%      & 417.5355              & 449.6927              \\
	&                                 &                                 &                             &                                 &                             &                                  & 90\%      & 573.6224              & 666.8224              \\ \hline
	\multirow{3}{*}{3LN}  & \multirow{3}{*}{12.40}          & \multirow{3}{*}{25.30}          & \multirow{3}{*}{5}          & \multirow{3}{*}{27.48}          & \multirow{3}{*}{4}          & \multirow{3}{*}{1.5978}          & 70\%      & 71.2881               & 135.4887              \\
	&                                 &                                 &                             &                                 &                             &                                  & 80\%      & 112.9413              & 139.0838              \\
	&                                 &                                 &                             &                                 &                             &                                  & 90\%      & 136.5332              & 144.6103              \\ \hline
	\multirow{3}{*}{4LN}  & \multirow{3}{*}{12.36}          & \multirow{3}{*}{25.26}          & \multirow{3}{*}{4}          & \multirow{3}{*}{27.65}          & \multirow{3}{*}{5}          & \multirow{3}{*}{1.5911}          & 70\%      & 25.3765               & 69.6743               \\
	&                                 &                                 &                             &                                 &                             &                                  & 80\%      & 25.9574               & 77.0592               \\
	&                                 &                                 &                             &                                 &                             &                                  & 90\%      & 66.5130               & 145.5590              \\ \hline
	\multirow{3}{*}{5LN}  & \multirow{3}{*}{12.36}          & \multirow{3}{*}{25.54}          & \multirow{3}{*}{7}          & \multirow{3}{*}{29.18}          & \multirow{3}{*}{7}          & \multirow{3}{*}{1.5898}          & 70\%      & 21.2581               & 51.4987               \\
	&                                 &                                 &                             &                                 &                             &                                  & 80\%      & 25.7930               & 66.4282               \\
	&                                 &                                 &                             &                                 &                             &                                  & 90\%      & 55.3524               & 113.0587              \\ \hline
	\multirow{3}{*}{6LN}  & \multirow{3}{*}{12.36}          & \multirow{3}{*}{25.72}          & \multirow{3}{*}{10}         & \multirow{3}{*}{30.07}          & \multirow{3}{*}{10}         & \multirow{3}{*}{1.5896}          & 70\%      & 25.9670               & 122.5587              \\
	&                                 &                                 &                             &                                 &                             &                                  & 80\%      & 34.3358               & 145.0191              \\
	&                                 &                                 &                             &                                 &                             &                                  & 90\%      & 71.3076               & 194.7760              \\ \hline
	\multirow{3}{*}{2GB2} & \multirow{3}{*}{12.41}          & \multirow{3}{*}{\textbf{25.15}} & \multirow{3}{*}{\textbf{1}} & \multirow{3}{*}{27.17}          & \multirow{3}{*}{2}          & \multirow{3}{*}{1.6096}          & 70\%      & 70.4405               & 76.1168               \\
	&                                 &                                 &                             &                                 &                             &                                  & 80\%      & 134.6649              & 130.9848              \\
	&                                 &                                 &                             &                                 &                             &                                  & 90\%      & 232.3097              & 198.0295              \\ \hline
	\multirow{3}{*}{3GB2} & \multirow{3}{*}{12.37}          & \multirow{3}{*}{25.24}          & \multirow{3}{*}{3}          & \multirow{3}{*}{27.42}          & \multirow{3}{*}{3}          & \multirow{3}{*}{\textbf{1.5328}} & 70\%      & 25.2301               & 49.2974               \\
	&                                 &                                 &                             &                                 &                             &                                  & 80\%      & 53.8051               & 72.5710               \\
	&                                 &                                 &                             &                                 &                             &                                  & 90\%      & 196.2506              & 193.4914              \\ \hline
	\multirow{3}{*}{4GB2} & \multirow{3}{*}{12.37}          & \multirow{3}{*}{25.40}          & \multirow{3}{*}{6}          & \multirow{3}{*}{28.31}          & \multirow{3}{*}{6}          & \multirow{3}{*}{1.5884}          & 70\%      & 16.7314               & 56.4453               \\
	&                                 &                                 &                             &                                 &                             &                                  & 80\%      & 43.0186               & 80.2198               \\
	&                                 &                                 &                             &                                 &                             &                                  & 90\%      & 209.3168              & 164.2237              \\ \hline
	\multirow{3}{*}{5GB2} & \multirow{3}{*}{12.36}          & \multirow{3}{*}{25.55}          & \multirow{3}{*}{8}          & \multirow{3}{*}{29.18}          & \multirow{3}{*}{7}          & \multirow{3}{*}{1.5857}          & 70\%      & \textbf{16.5706}      & \textbf{26.7718}      \\
	&                                 &                                 &                             &                                 &                             &                                  & 80\%      & \textbf{32.3473}      & \textbf{33.4173}      \\
	&                                 &                                 &                             &                                 &                             &                                  & 90\%      & \textbf{51.3189}      & \textbf{63.0882}      \\ \hline
	\multirow{3}{*}{6GB2} & \multirow{3}{*}{\textbf{12.35}} & \multirow{3}{*}{25.71}          & \multirow{3}{*}{9}          & \multirow{3}{*}{30.06}          & \multirow{3}{*}{9}          & \multirow{3}{*}{1.5830}          & 70\%      & 55.8262               & 97.3636               \\
	&                                 &                                 &                             &                                 &                             &                                  & 80\%      & 66.3440               & 129.4636              \\
	&                                 &                                 &                             &                                 &                             &                                  & 90\%      & 97.5921               & 231.7915              \\ \hline
\end{tabular}
	\label{tab:emLDMM_gb2_performance}
\end{table*}

\subsubsection{Posterior distribution of parameters}

We consider the mixture of log-normal distribution and  Pareto distribution in this section.
We assume  that the conjugate prior  for the log-normal likelihood is 
$\mu, \sigma^2\sim N\mbox{-}\Gamma^{-1}(8, 100, 1, 1),$
and the conjugate prior distribution for the Pareto likelihood with known minimum $\sigma$ is 
{$\alpha\sim \Gamma(25000, 50000).$}
Please refer to Appendix \ref{appendix:loss} for the technical details of those distributions.
The convergence of the Gibbs sampling for Model 1LN1P is shown in the Figure \ref{fig:lp}. 
%\blue{To obtain appropriate initial values for Gibbs sampling, we first extract the distribution parameters from the posterior conjugate distribution based on random categorical data.} 
%Then we reassign a cluster for each case $i$ according to the conditional distribution $p(y_i=k|\vec y_{\neg i},\vec d, \vec x)$ and sample loss distribution parameters from the posterior conjugate distributions. 

\begin{figure}[h!]

\centering
        \includegraphics[width=0.49\linewidth]{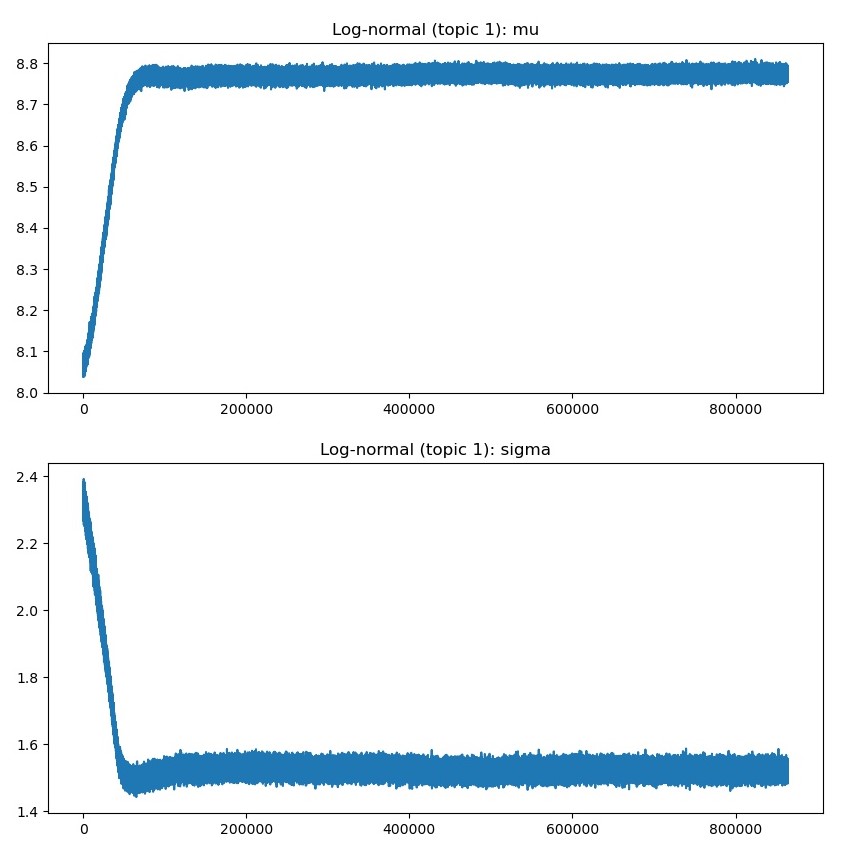}
        \includegraphics[width=0.49\linewidth]{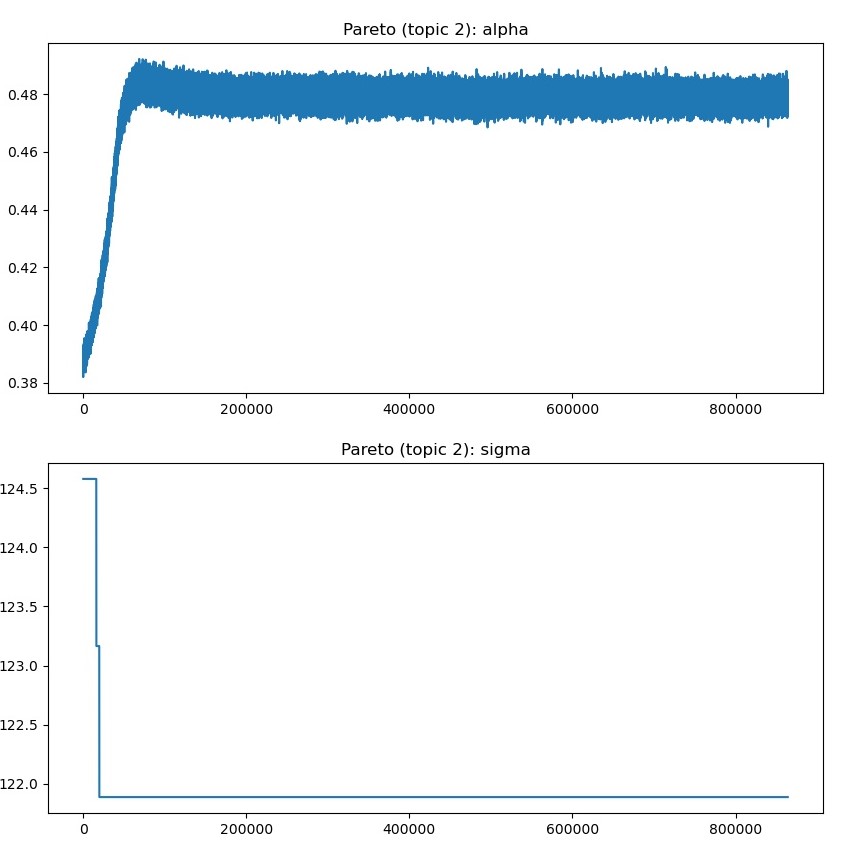}
\caption{The convergence of Algorithm \ref{alg:gs} for Model 1LN1P.}
\label{fig:lp}%文中引用该图片代号
\end{figure}

Under the condition of $K\le3$, Table \ref{tab:gsLDMM_estimation2} present the estimated parameters for all the possible LDMM models: 2LN, 2P, 1LN1P, 3LN, 3P, 2LN1P, 1LN2P.
Similar to Table \ref{tab:emLDMM_GB2_parameter}, It shows that different component loss distributions have quite different parameters.
The top-5 words with the highest occurrence probabilities in each topic somehow explain the characteristics of each  component. 
The components with higher means are related to  severe claims which are normally featured by the keywords {\it back, shoulder}.
In contrast, the components with lower means are related to mild claims which are normally featured by the keywords {\it finger, eye}.
This aligns with the observations in Table \ref{tab:emLDMM_GB2_parameter}.

\begin{table*}[h!]
    \small
    \renewcommand\arraystretch{1.2}
    \centering
    \caption{The  estimated posterior mean of parameters and the top-5 words in the LDMM models with component loss of LN or P under $K\le 3$.}
    \begin{tabular}{ccccccc}
    \hline\hline
    Models & $\theta$ & Components & Posterior mean & Posterior mean  & Top-5 words\\
    \hline
    \multirow{2}*{2LN} & 0.38 & Log-normal&$\mu = 7.24$&$\sigma =1.67$&finger, hand, struck, laceration, lacerate\\
    &0.41& Log-normal &$\mu=8.56$&$\sigma=2.07$&back, strain, lifting, shoulder, eye\\\hline
    \multirow{2}*{2P} & 0.43 & Pareto&$\sigma = 122$&$\alpha =0.47$&finger, hand, struck, eye, laceration\\
    &0.57& Pareto &$\sigma=219$&$\sigma=0.37$& back, strain, lifting, shoulder, fell\\\hline
    \multirow{2}*{1LN1P} & 0.59 & Log-normal&$\mu =8.78$&$\sigma =1.52$&back, strain, lifting, shoulder, fell\\
    &0.41& Pareto &$\sigma=0.48$&$\alpha=121.89$&finger, hand, struck, laceration, eye\\\hline
    \multirow{3}*{3LN} &0.30 & Log-normal&$\mu = 9.08$&$\sigma =1.13$&back, strain, lifting, shoulder, strained\\
    &0.35 & Log-normal&$\mu = 6.89$&$\sigma =1.45$&finger, hand, struck, laceration, lacerated\\
    &0.35 & Log-normal&$\mu = 8.37$&$\sigma =1.86$&eye, foreign, body, grinding, burn\\\hline
    \multirow{3}*{3P} & 0.35 & Pareto&$\sigma = 140$&$\alpha =0.46$&finger, hand, struck, laceration, lacerated\\
    &0.12 & Pareto&$\sigma = 122$&$\alpha =0.49$&eye, foreign, body, grinding, burn\\
    &0.53 & Pareto&$\sigma = 240$&$\alpha =0.38$&back, strain, lifting, shoulder, knee\\\hline
    \multirow{3}*{2LN1P} & 0.35 & Log-normal&$\mu = 9.02$&$\sigma =1.30$&back, strain, lifting, shoulder, strained\\
    &0.34 & Log-normal&$\mu = 8.18$&$\sigma =1.73$&fell, knee, slipped, ankle, injury\\
    &0.31 & Pareto&$\sigma = 122$&$\alpha =0.50$&finger, eye, struck, hand, laceration\\\hline
    \multirow{3}*{1LN2P} & 0.56 & Log-normal&$\mu = 8.79$&$\sigma =1.49$&back, strain, lifting, shoulder, fell\\
    &0.36 & Pareto&$\sigma = 121$&$\alpha =0.48$&finger, struck, hand, laceration, eye\\
    &0.01 & Pareto&$\sigma = 148$&$\alpha =0.47$&accident, vehicle, burn, motor, hand\\
    \hline\hline
    \end{tabular}
    \label{tab:gsLDMM_estimation2}
\end{table*}

We systematically compare the LDMM models under different number of components $K=1,\ldots,4$. 
Table \ref{tab:gsLDMM_performance2} compares all the possible combinations in terms of the NLL, AIC, BIC and DIC. 
{Note that the AIC and BIC are calculated at the MAP estimates, while the DIC is calculated in the MCMC simulations.}
For each $K=1,2,3,4$, the models with the smallest DIC are bolded. 
% It should be noted that the fitting performance of the model is closely related to the choice of hyperparameters of the parameter prior distribution; selecting different hyperparameters will yield different fitting results. 
It seems that $K=3,4$ components are sufficient to capture the complexity in the claims amount distribution.

\begin{table*}[h!]
\small
    \centering
    \caption{Comparison of the LDMM models with all the possible combinations of $LN$ and $P$ under $K=1,\ldots,4$.}
    \begin{tabular}{cccccc}
    \hline\hline
    Models &NLL & AIC & BIC &DIC& DIC Rank\\
    \hline
    \multicolumn{6}{c}{Single-component models}\\
    \hline
    \textbf{1LN} & \textbf{37.96} &\textbf{ 76.10} &\textbf{76.83} &\textbf{75.90}&\textbf{13}\\
    1P & 38.31 & 76.80 &74.53 &76.63&14\\
    \hline
    \multicolumn{6}{c}{Two-component models}\\
    \hline
    2LN& 36.70&73.74 &74.29&73.46&11\\
    \textbf{1LN1P} &\textbf{36.49} & \textbf{73.32} & \textbf{73.88} &\textbf{72.99}&\textbf{10}\\
    2P & 36.83 & 73.99 &74.54 &73.65&12\\
    \hline
    \multicolumn{6}{c}{Three-component models}\\
    \hline
    \textbf{3LN} &\textbf{35.80} & \textbf{72.09} & \textbf{72.49} &\textbf{71.59}&\textbf{6}\\
    2LN1P & 35.86& 72.22& 72.61&71.72&8\\
    1LN2P & 35.80&72.11 & 72.50 &71.60&7\\
    3P & 36.16 & 72.81 & 73.20 &72.31&9\\
    \hline
    \multicolumn{6}{c}{Four-component models}\\
    \hline
    \textbf{4LN} &\textbf{35.28} & \textbf{71.22} & \textbf{71.45} &\textbf{70.55}&\textbf{1}\\
    3LN1P &35.61&71.90& 72.12&71.72&4\\
    2LN2P &35.30&71.27 & 71.50 &71.22&3\\
    1LN3P &35.55&71.76 & 71.99 &71.10&2\\
    4P &35.71 & 72.10 & 72.32 &71.43&5\\
    \hline\hline
    \end{tabular}
    \label{tab:gsLDMM_performance2}
\end{table*}

The selected Models 1LN, 1LN1P, 3LN and 4LN are investigated further in Table \ref{tab:gsLDMM_result}, where the NLL, AIC, BIC, DIC , perplexity and Wasserstein distance are calculated.
Model 4LN with the smallest DIC also has the lowest perplexity and the lowest Wassertein distance.
Roughly speaking, Model 4LN has the best performance from either the DMM model or  the finite mixture model perspective. 
%Moreover, it is noteworthy that under the prior assumptions of this paper, the LP introduced the Pareto distribution, but this did not improve the fit on the loss distribution; on the contrary, it deteriorated. 
Finally, we check the stability of the sampled parameters in Model 4LN. The results are presented  in Table \ref{tab:lwwptop20_stablility} in Appendix \ref{appendix:stability}.
The word occurrence probabilities $\bphi_k^{[t]}$ in each topic $k$ maintains good stability during the MCMC sampling process $t=1,\ldots,T$.

%In summary, the fitting performance of the EM-LDMM algorithm is much better than that of the GS-LDMM algorithm, which is due to the Gibbs sampling algorithm converging to the local optimal solution. Furthermore, we found that by introducing Log-normal, Weibull, and Pareto distributions, the claim characteristics of different types of claim events can be well distinguished. In addition, by introducing a more flexible GB2 distribution family, we can solve the LDMM problem while ensuring better fitting of the loss distribution, providing a theoretical basis for risk assessment and classification of claim events. Last but not least, by observing the high-frequency words in Table \ref{tab:gsLDMM_estimation} and comparing the perplexity in Tables \ref{tab:emLDMM_gb2_performance} and \ref{tab:gsLDMM_result}, it can be found that the model obtained by GS-LDMM has smaller perplexity, which can better identify different types of claim events.

\begin{table*}[h!]
    \small
    \centering
    \caption{Comparison of the 4 selected LDMM models.}
    \begin{tabular}{cccccccccrrr}
    \hline\hline
    \multirow{2}*{Models} & \multirow{2}*{NLL} & \multirow{2}*{AIC} & \multirow{2}*{BIC} & \multirow{2}*{DIC} & \multirow{2}*{DIC Rank} & \multirow{2}*{perplexity} & \multicolumn{3}{c}{Wassertein distance}\\
    \cline{8-10}
    &&&&&&&percent&training  & test \\
    \hline
    \multirow{3}*{LN} & \multirow{3}*{37.96}&\multirow{3}*{76.10}&\multirow{3}*{76.83}&\multirow{3}*{75.90}&\multirow{3}*{13}&\multirow{3}*{1.0006}&70\%&186.4453& 	207.3149\\
    &&&&&&&80\%&260.2249 &254.8153 \\
    &&&&&&&90\%&306.2297&291.5198 \\
    \hline
    \multirow{3}*{1LN1P} & \multirow{3}*{36.49}&\multirow{3}*{73.32}&\multirow{3}*{73.88}&\multirow{3}*{72.99}&\multirow{3}*{10}&\multirow{3}*{1.0005}&70\%&248.4216& 	214.8590\\
    &&&&&&&80\%&558.7675 &507.4566 \\
    &&&&&&&90\%&928.7387&920.2910\\
    \hline
    \multirow{3}*{3LN} & \multirow{3}*{35.80}&\multirow{3}*{72.09}&\multirow{3}*{72.49}&\multirow{3}*{71.59}&\multirow{3}*{6}&\multirow{3}*{1.0004}&70\%&157.5479& 	 159.6223\\
    &&&&&&&80\%&269.6857& 	 286.6179\\
    &&&&&&&90\%&343.7362& 	 351.1413\\
    \hline
    \multirow{3}*{4LN} & \multirow{3}*{35.28}&\multirow{3}*{71.22}&\multirow{3}*{71.45} &\multirow{3}*{70.55}&\multirow{3}*{1}&\multirow{3}*{\textbf{1.0003}}&70\%& \textbf{148.2713}& 	\textbf{143.3812} \\
    &&&&&&&80\%&\textbf{263.9898}& 	\textbf{259.3970} \\
    &&&&&&&90\%&\textbf{346.1372}& 	\textbf{327.4872} \\
    \hline\hline
    \end{tabular}
    \label{tab:gsLDMM_result}
\end{table*}

\subsection{Prediction on the test RBNS claims}

We predict the claims amount on the test claims, which are reported to the insurance companies but not settled (called RBNS claims).
We first simulate the parameters by Algorithm \ref{alg:gs}, then simulate the claims amount for the RBNS claims by Algorithm \ref{alg:risk_sampling}.
Once a sample of simulated claims amount is obtained, the VaR and CTE can be estimated by equations \eqref{var} and \eqref{cte}.
Tables \ref{tab:2ln_risk} and \ref{tab:llll_risk}  present the estimated VaR and CTE for a selection of RBNS claims from the test data under the 2LN or  4LN assumption. 
The first column shows the claim description text, the second column shows the true claims amount,  and 
the last two columns show the estimated VaR and CTE at the confidence levels of 95\%.

\begin{table*}[h!]
	\renewcommand\arraystretch{1.2}
	\small
	\centering
	\caption{The predicted risk measures under the 2LN assumption.}
	\begin{tabular}{cccccc}
		\hline\hline
		\multirow{2}*{Claim description} & \multirow{2}*{True loss} &95\% & 95\% \\ 
		\cline{3-4} 
		&&VaR&CTE \\
		\hline
		stepped forklift twisted knee wrist 	&5767.90&	15581.56	&22679.03	\\
		struck piece metal eye& 	500.10	&15635.55	&23999.30\\
		struck knife cut index finger &	579.20	&15580.84&	22454.22	\\
		back lifting rubbish bin back injury&	19544.8&		15460.73&	22489.52\\
		\hline\hline
	\end{tabular}
	\label{tab:2ln_risk}
\end{table*}

\begin{table*}[h!]
	\renewcommand\arraystretch{1.2}
	\small
	\centering
	\caption{The predicted risk measures under the 4LN assumption.
	}
	\begin{tabular}{ccccccc}
		\hline\hline
		\multirow{2}*{Claim description} & \multirow{2}*{True loss} &95\% &95\% \\ 
		\cline{3-4}
		&&VaR&CTE \\
		\hline
		stepped crates truck tray fracture forearm &6326.30&80513.22&249746.40 \\
		cut sharp edge cut thumb&2293.90&8531.00&18548.10 \\
		strained muscle back strained back pain &4464.00&58364.70&101999.50 \\
		struck hot water burn hand shoulder&1888.40&63606.80&201930.60 \\
		\hline\hline
	\end{tabular}
	\label{tab:llll_risk}
\end{table*}

Finally, we calculate the percentages of the true loss smaller than the predicted VaR on the test data for Models 1LN, 1LN1P, 3LN and 4LN as follows:
$$\frac{\sum_{i\in \mathcal{I}_{test}}\mathbb{I}(Y_{i}<VaR_i(\alpha))}{|\mathcal{I}_{test}|},$$
where $\mathbb{I}$ is an indicator function and the denominator is the sample size of test data.
The results are presented in Table \ref{tab:all_risk}. 
It shows that the proposed models provide a good estimate of VaR for the test dataset.

\begin{table*}[h!]
	\renewcommand\arraystretch{1.2}
	\small
	\centering
	\caption{The accuracy of the estimated VaR on the test data.}
	\begin{tabular}{cccc}
		\hline\hline
		{Models}   &95\% VaR &99\% VaR \\ 
		\hline
		1LN &  99.19\% &100.00\% \\
		1LN1P &  97.27\% &99.63\% \\
		3LN &  95.61\% &99.27\% \\
		4LN& 95.32\%  &98.95\%\\
		\hline\hline
	\end{tabular}
	\label{tab:all_risk}
\end{table*}

\subsection{Component analysis}

In this section, we interpret the loss components in Models 2LN, 2GB2 and 4LN.
Normally, it is difficult to explain the loss components in the finite mixture models.
However, in our proposed LDMM model the loss component can be explained by the corresponding claim description topic.
As seen in Tables \ref{tab:emLDMM_GB2_parameter} and \ref{tab:gsLDMM_estimation2}, the most probable words in each topic can somehow explains why a claim belongs to a loss component.

Figures \ref{fig:2ln2} and   \ref{fig:2gb2_2} compare the posterior predictive  distribution of the claim amount (solid line) with the empirical distribution of the true claim amount (bars) for the test dataset under the assumption of 2LN and 2GB2, respectively. 
The left panel shows the distributions below $20,000$ and the right panel shows the distribution above $20,000$.
The associated tables list the proportion  below/above $20,000$ and the summary statistics of the true claim amounts in each component, respectively.
It is worth noting that the LDMM model can decompose the empirical loss distribution very well. 
The claims in the first component/topic are more severe than those in the second component/topic. 
The claims in the first component/topic exhibits characteristics of large mean and thick tail with higher reserving risk, 
while the claims in the second component/topic exhibits characteristics of small mean and thin tail with lower reserving risk.

%\begin{figure}[h!]
%    \begin{minipage}{1.05\linewidth}
%		%\centering
%		\includegraphics[width=0.92\linewidth]{fig/component_analysis/2ln_fitted.png}
%	\end{minipage}
%    \renewcommand\arraystretch{1.1}\small
%    %\centering
%    \begin{subtable}
%    \small
%    \centering
%        \begin{tabular}{lrccccc}
%            \hline\hline
%            k&&Min&25\%&Median&75\%&Max\\\hline
%            1&81\%&816&3495&5643&8355&19999 \\
%            2&100\%& 121&412&682&1135&4501\\
%    \hline\hline
%        \end{tabular}
%    \end{subtable}
%    \hfill\qquad
%    \begin{subtable}
%    \small
%        \begin{tabular}{lrccccc}
%            \hline\hline
%            k&&Min&25\%&Median&75\%&Max\\\hline
%            1&19\%&20004&26391&38078&68787&4027136\\
%            2&0\%& ---&---&---&---&---\\
%    \hline\hline
%        \end{tabular}
%    \end{subtable}
% \caption{Loss fitting effect of each topic on training samples (2LN). \blue{Legend should be loss.}}	\label{fig:2ln1}%文中引用该图片代号
% \end{figure}

 \begin{figure}[h!]
 	
    \begin{minipage}{\linewidth}
		\centering
		\includegraphics[width=0.95\linewidth]{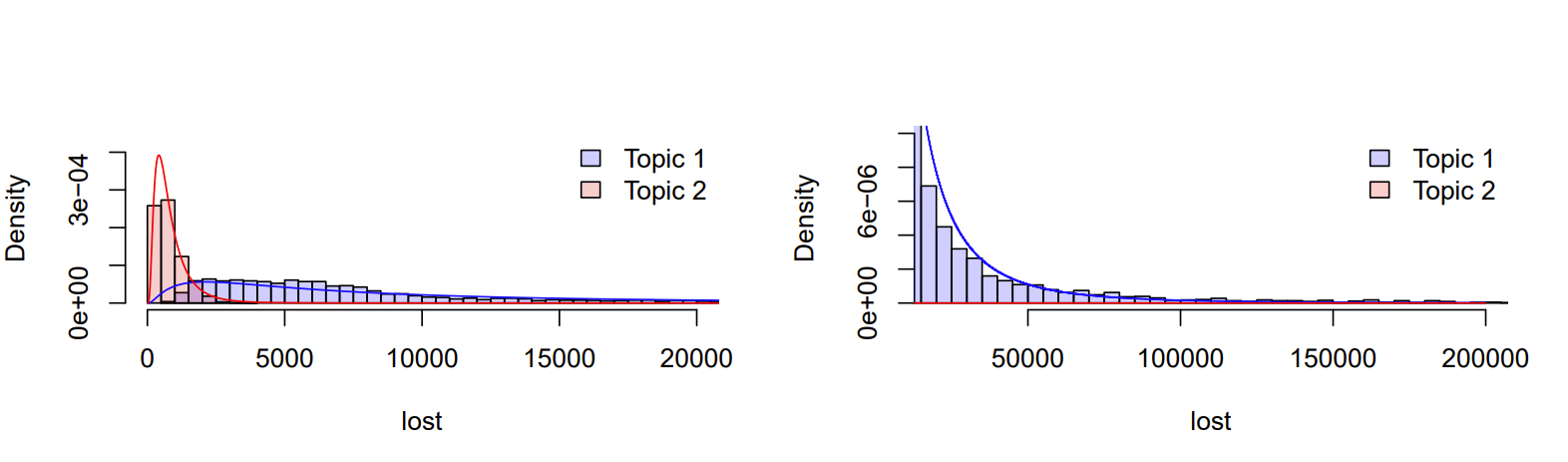}

	\end{minipage}
\small
    \begin{subtable}
    \small
    \centering
        \begin{tabular}{ccccccc}
            \hline\hline
            $k$& Prop. &Min&25\%&Median&75\%&Max\\\hline
            1&82\%&629&$3,370$&$5,640$&$8,368$&$19,957$ \\
            2&100\%& 123&407&662&$1,067$&$3,656$\\
    \hline\hline
        \end{tabular}
    \end{subtable}
    \hfill\qquad
    \begin{subtable}
    \small
    \centering
        \begin{tabular}{ccccccc}
            \hline\hline
            $k$& Prop. &Min&25\%&Median&75\%&Max\\\hline
            1&18\%&$20,023$&$26,824$&$38,620$&$68,552$&$742,003$\\
            2&0\%& ---&---&---&---&---\\
    \hline\hline
        \end{tabular}
    \end{subtable}
   \caption{The posterior predictive distribution of the claim amount (solid line) and the empirical distribution of the true claim amount (bars) for the test data under the assumption of 2LN.}		\label{fig:2ln2}%文中引用该图片代号
 \end{figure}

%\begin{figure}[h!]
%      
%    \begin{minipage}{1.05\linewidth}
%		%\centering
%		\includegraphics[width=0.92\linewidth]{fig/component_analysis/2gb2fitall1.png}
%
%	\end{minipage}
%    \renewcommand\arraystretch{1.1}\small
%    %\centering
%    \begin{subtable}
%    \small
%    \centering
%        \begin{tabular}{lrccccc}
%            \hline\hline
%            k&&Min&25\%&Median&75\%&Max\\\hline
%            1&81\%&816&3495&5643&8355&19999 \\
%            2&100\%& 121&412&682&1135&4501\\
%    \hline\hline
%        \end{tabular}
%    \end{subtable}
%    \hfill\qquad
%    \begin{subtable}
%    \small
%        \begin{tabular}{lrccccc}
%            \hline\hline
%            k&&Min&25\%&Median&75\%&Max\\\hline
%            1&19\%&20004&26391&38078&68787&4027136\\
%            2&0\%& ---&---&---&---&---\\
%    \hline\hline
%        \end{tabular}
%    \end{subtable}
%  \caption{Loss fitting effect of each topic on training samples (2GP2)}		\label{fig:2gb2_1}%文中引用该图片代号
% \end{figure}

\begin{figure}[h!]
    \begin{minipage}{\linewidth}
		%\centering
		\includegraphics[width=0.95\linewidth]{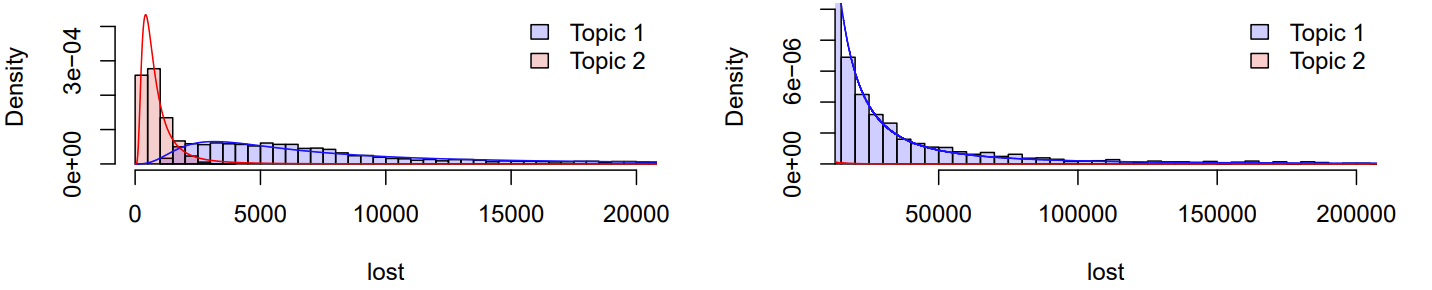}
	\end{minipage}
    \renewcommand\arraystretch{1}\small
    %\centering
    \begin{subtable}
    \small
    \centering
        \begin{tabular}{ccccccc}
            \hline\hline
            $k$& Prop. &Min&25\%&Median&75\%&Max\\\hline
            1&81\%& $1,002$ & $3,589$ & $5,768$ & $8,519$ & $19,957$\\
            2&100\%& 123&416&685& $1,124$& $3,656$\\
    \hline\hline
        \end{tabular}
    \end{subtable}
    \hfill\qquad
    \begin{subtable}
    \small
        \begin{tabular}{ccccccc}
            \hline\hline
            $k$& Prop. &Min&25\%&Median&75\%&Max\\\hline
            1&19\%&$20,022$&$26,824$&$38,619$&$68,552$&$742,003$\\
            2&0\%& ---&---&---&---&---\\
    \hline\hline
        \end{tabular}
    \end{subtable}
\caption{The posterior predictive distribution of the claim amount (solid line) and the empirical distribution of the true claim amount (bars) for the test data under the assumption of 2GB2.}		\label{fig:2gb2_2}%文中引用该图片代号
\end{figure}

Table \ref{tab:2ln2gb2_top_20_words} shows the top 20 words with the highest occurrence probabilities in each component/topic. 
The first component loss distribution has a higher mean and risk measures with higher reserving risk compared to the second one. 
The claims in the first component are mainly due to injuries to  the back, shoulder and knees caused by lifting heavy objects.
In contrast, the claims in the second component are mainly due to  hand cuts and eye injuries.
In summary, Models 2LN and 2GB2 are very similar and can split claims  into two  components/topics with significant different claims amounts and claims features. 

\begin{table*}[h!]
	\small
	\renewcommand\arraystretch{1.3}
	\centering
	\caption{The top-20 words in each component/topic in Models 2LN and 2GB2.}
	\begin{tabular}{ccccc}
		\hline\hline
		Model & Topic &  Component  & Distribution  & TOP-20 words\\
		\hline
		\multirow{8}*{2LN} & \multirow{4}*{1} & \multirow{4}*{Log-normal} & \multirow{4}*{\includegraphics[width=0.25\linewidth]{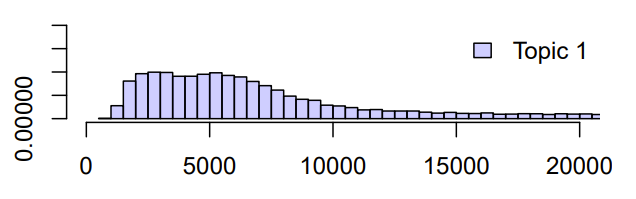}}& back, strain, lifting, shoulder, fell, \\ 
		&&&&knee, strained, injury, slipped, neck, \\
		&&&&hand, finger, wrist, ankle, struck, arm, \\
		&&&&pain, foot, hit, laceration\\
		\cline{2-5}
		&\multirow{4}*{2} & \multirow{4}*{Log-normal} &\multirow{4}*{\includegraphics[width=0.25\linewidth]{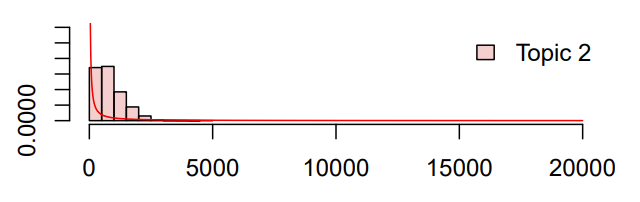}}& finger, struck, hand, eye, laceration, \\
		&&&&foreign, body, lacerated, thumb, cut, \\
		&&&&knife, index, slipped, metal, bruised, \\
		&&&&fell, hit, wrist, foot, caught\\
		\hline
		\multirow{8}*{2GB2} &\multirow{4}*{1} & \multirow{4}*{GB2}&\multirow{4}*{\includegraphics[width=0.25\linewidth]{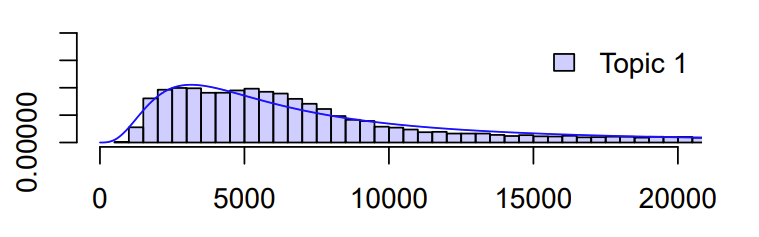}}& back, stain, lifting, shoulder, fell, \\
		&&&&knee, strained, injury, slipped, neck, \\
		&&&&hand, finger, wrist, ankle, struck, \\
		&&&&pain, arm, foot, hit, soft\\
		\cline{2-5}
		&\multirow{4}*{2} & \multirow{4}*{GB2} &\multirow{4}*{\includegraphics[width=0.25\linewidth]{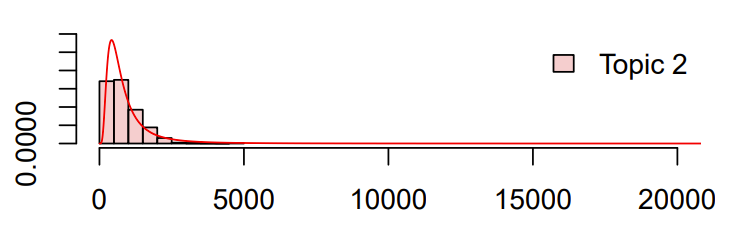}}&finger, struck, hand, eye, laceration, \\
		&&&& lacerated, foreign, body, thumb, cut, \\
		&&&& knife, slipped, index, bruised, metal, \\
		&&&& fell, hit, wrist, foot, caught\\
		\hline\hline
	\end{tabular}
	\label{tab:2ln2gb2_top_20_words}
	
\end{table*}

%\begin{figure}[h!]
%    \begin{minipage}{1.05\linewidth}
%    \includegraphics[width=0.92\linewidth]{fig/component_analysis/llllfitall1.png}
%    \end{minipage}
%    \renewcommand\arraystretch{1.1}\small
%    \centering
%    \begin{subtable}
%        \centering
%        \scriptsize
%        \begin{tabular}{lrlccccc}
%            \hline\hline
%            k&&Min&25\%&Median&75\%&Max\\\hline
%            1&80\%&366&4388&6287&8668&19999\\
%            2&84\%& 227&1253&2873&6337&19989\\
%            3&87\%& 233&1212&2716&5775&19996\\
%            4&97\%&121&378&657&1455&19947 \\
%    \hline\hline
%        \end{tabular}
%    \end{subtable}
%    \hfill\qquad
%    \begin{subtable}
%    \small
%        \begin{tabular}{lrlccccc}
%            \hline\hline
%            k&&Min&25\%&Median&75\%&Max\\\hline
%            1&20\%&20004&26636&39387&76914&865770\\
%            2&16\%& 20065&28648&40741&65489&469915\\
%            3&13\%& 20006&26444&37995&63748&823706\\
%            4&3\%&20042&25533&25533&60837&558409   \\
%    \hline\hline
%        \end{tabular}
%    \end{subtable}
%  \caption{Loss fitting effect on training samples (LLLL) using GS-LDMM method \blue{train-test?}}    \label{fig:llll1}%文中引用该图片代号
%\end{figure}

Table \ref{tab:gsLDMM_result} shows that Model 4LN performs the best in terms of all the metrics.
So we analyze the resulted four components in Model 4LN.
Figure \ref{fig:llll2}  compares the posterior predictive  distribution of the claim amount (solid line, estimated from Model 4LN) with the empirical distribution of the true claim amount (bars) for the test dataset. 
The left panel shows the distributions below $20,000$ and the right panel shows the distribution above $20,000$.
The associated tables list the proportion  below/above $20,000$ and the summary statistics of the true claim amounts in each component, respectively.
The third component has the highest mean and the thickest tail with the highest reserving risk,
while the  fourth component has the lowest mean and the  thinnest tail with the lowest reserving risk. 

Table \ref{tab:llll_top_20_words} shows the top 20 words with the highest occurrence probabilities for each component/topic.
The claims in the third component have the highest reserving risk, which involve injuries to the {\it knee, ankle, foot, wrist}, etc.
The claims in the fourth component have the lowest reserving risk, which are mainly caused by burn or foreign objects entering the eyes.
It is interesting that  the third and fourth components are quite different in terms of the top 10 words, which indicates that we can classify the claims into those two categories with a high degree of confidence.

\begin{figure}[h!]
	\begin{minipage}{\linewidth}
		\includegraphics[width=0.99\linewidth]{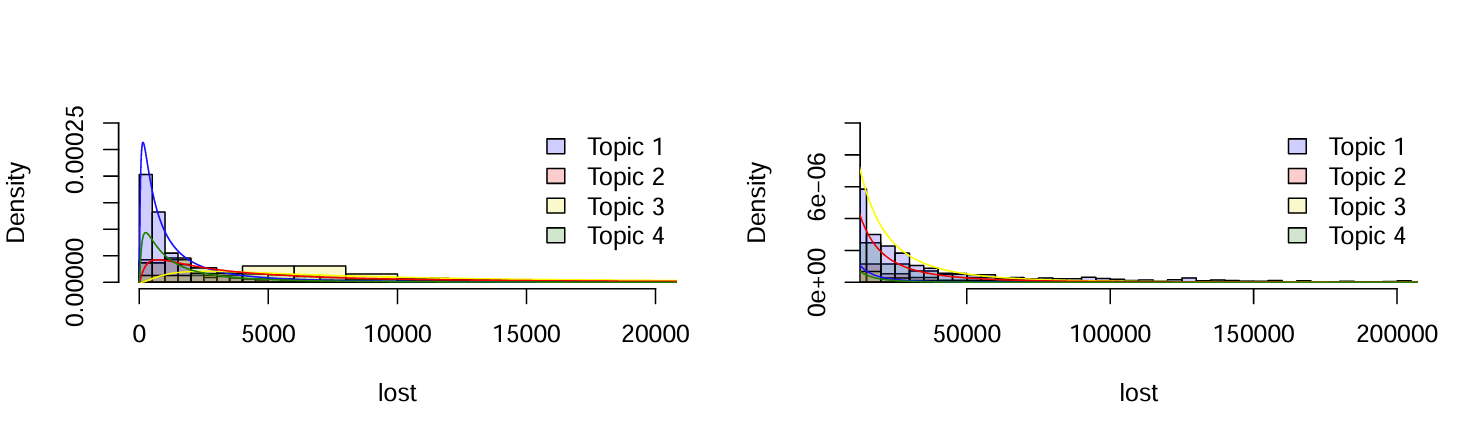}
	\end{minipage}
	\renewcommand\arraystretch{1}\small
	%\centering
	\begin{subtable}
		\footnotesize
		\centering
		\begin{tabular}{ccccccc}
			\hline\hline
			$k$&Prop.&Min&25\%&Median&75\%&Max\\\hline
			1&80\%& 124&814&$2,381$&$6,064$&$19,946$ \\
			2&90\%& 123&741&$2,200$&$6,193$&$19,819$\\
			3&86\%& 142&787&$2,617$&$6,463$&$19,925$\\
			4&97\%& 147&830&$2,745$&$6,458$&$19,957$\\
			\hline\hline
		\end{tabular}
	\end{subtable}
	\hfill\qquad
	\begin{subtable}
		\small
		\begin{tabular}{ccccccc}
			\hline\hline
			$k$&Prop.&Min&25\%&Median&75\%&Max\\\hline
			1&20\%&$20,063$&$26,592$&$35,919$&$68,708$&$608,650$  \\
			2&10\%&$20,022$&$26,763$&$41,971$&$72,181$&$742,003$\\
			3&14\%& $20,109$&$28,722$&$38,871$&$66,135$&$501,899$\\
			4&3\%& $20,065$&$25,837$&$38,693$&$64,708$&$374,588$\\
			\hline\hline
		\end{tabular}
	\end{subtable}
	\caption{The posterior predictive distribution of the claim amount (solid line) and the empirical distribution of the true claim amount (bars) for the test data under the assumption of 4LN.} \label{fig:llll2}%文中引用该图片代号
\end{figure}

\begin{table*}[h!]
    \small
    \renewcommand\arraystretch{1.3}
    \centering
    \caption{The top-20 words in each component/topic in Model 4LN.}
    \begin{tabular}{ccccc}
    \hline\hline
    Topic &  Component  & Distribution  & Top-20 words\\
    \hline
    \multirow{4}*{1} & \multirow{4}*{Log-normal} &\multirow{4}*{\includegraphics[width=0.25\linewidth]{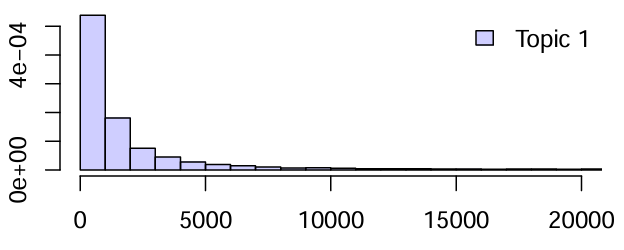}}& back, strain, lifting, shoulder, strained,\\ 
    &&&neck, pain, injury, slipped, accident,\\
    &&&vehicle, fell, motor, muscle, knee, upper\\
    &&&arm, repetitive, twisted\\
    \hline
    \multirow{4}*{2} & \multirow{4}*{Log-normal} &\multirow{4}*{\includegraphics[width=0.25\linewidth]{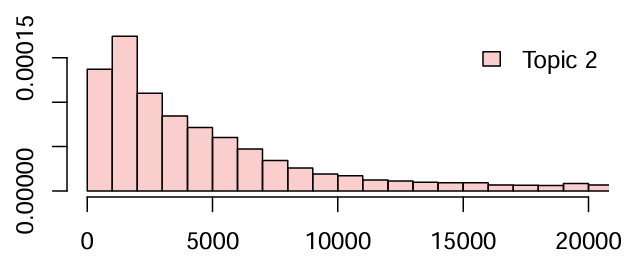}}& finger, hand, laceration, struck, lacerated,\\
    &&&cut, knife, thumb, index, slipped, caught\\
    &&&metal, hit, bruised, fell, puncture, ring\\
    &&&leg, arm, steel\\
    \hline
    \multirow{4}*{3} & \multirow{4}*{Log-normal}
    &\multirow{4}*{\includegraphics[width=0.25\linewidth]{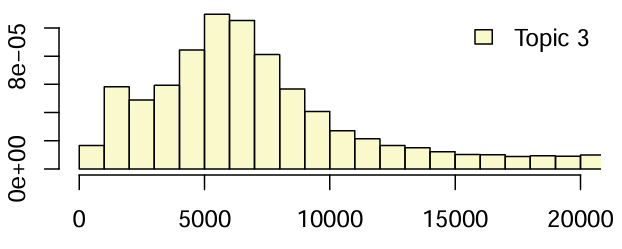}}& fell, knee, ankle, slipped, injury, bruised \\
    &&&wrist, foot, sprained, soft, tissue, back, \\
    &&&hit, elbow, shoulder,  strain, hand, arm\\
    &&&struck, strained\\
    \hline
    \multirow{4}*{4} & \multirow{4}*{Log-normal}&\multirow{4}*{\includegraphics[width=0.25\linewidth]{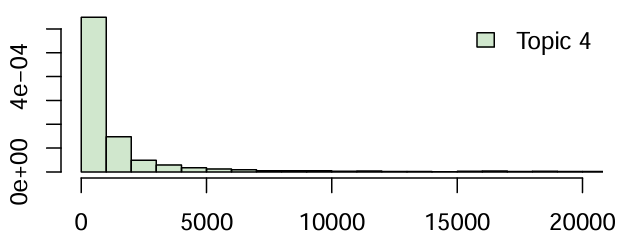}}&eye, foreign, body, grinding, burn, metal, \\
    &&&struck, wrist, hand, cornea, hot, foot, \\
    &&& tendon, steel, cleaning, synovitis, stress, \\
    &&& dust, shoulder, particle\\
    \hline\hline
    \end{tabular}
    \label{tab:llll_top_20_words}
    
\end{table*}

\section{Conclusions}

In this paper, we have successfully developed and validated a Loss Dirichlet Multinomial Mixture (LDMM) model, which represents an advancement in individual claims reserving using textual claims description. 
The LDMM model integrates an innovative Bayesian probabilistic framework to investigate the intricate relationship between claim amounts and their corresponding descriptions. 
By unifying these elements within a single Bayesian probabilistic model, the LDMM provides an insight of how claims amount and claims description interact.

Our empirical analysis reveals that the LDMM model is particularly adept at capturing complex characteristics of claims amount, including their multimodality, skewness, and heavy-tailed nature. This ability to model the inherent heterogeneity in loss data allows for a more nuanced understanding of the variability and patterns within the claims, which is very helpful for the RBNS claims reserving.

Moreover, the integration of claim descriptions into the model not only improves the fit of the model to the claims amount but also significantly boosts its explanatory power and practical utility. 
This enhancement is crucial in the context of the growing amount of unstructured data within the insurance industry. 
By leveraging the insights and methodologies developed through this research, the insurance sector can advance its digital transformation efforts, making more informed decisions and improving overall operational efficiency.

\section*{Funding}
Guangyuan Gao gratefully acknowledges the financial support from the MOE Project of Key Research Institute of Humanities and Social Sciences (22JJD910003). 
Yanxi Hou gratefully acknowledges the financial support from the National Natural Science Foundation of China (72171055).

\appendix

\section{Distributions used}\label{appendix:loss}

In the following, we list the distributions used in this paper. 
The Dirichlet distribution is the conjugate prior for the discrete distribution likelihood.
The normal-inverse gamma distribution is the conjugate prior for the log-normal loss distribution likelihood.
The log-normal, GB2 and Pareto distribution are used to model the claims amount.

\begin{enumerate}
	
	\item The PDF of a Dirichlet distribution with $d$-dimensional parameter vector $\balpha$ is as follows:
	\begin{equation}
		\mbox{Dir}(\bx; \balpha) = \frac{1}{C(\balpha)}\prod_{i=1}^d x_i^{\alpha_i-1}, 
	\end{equation}
where  $\bx=(x_1,\ldots,x_d)\in\mathbb (0,1)^d$ with the normalization $x_1+\cdots+x_d=1$  and
	\begin{equation}
	C(\balpha)=\frac{\prod_{i=1}^d\Gamma(\alpha_i)}{\Gamma(\sum_{i=1}^d\alpha_i)},  \quad d\ge 3.
\end{equation}
Here $C(\balpha)$ is the multivariate beta function and $\Gamma(\balpha)$ is the gamma function, a generalization of the factorial function. 
	Note that the Dirichlet distribution is an exponential family distribution.

\item The PDF of a normal-inverse gamma distribution with parameters $\mu_0,r,a,b$ is given by the product of a normal PDF and a gamma PDF:
$$f_{N-\Gamma^{-1}}(\mu,\sigma; \mu_0,r,a,b)=f_N(\mu;\mu_0,\sigma^2/r)f_{\Gamma^{-1}}(\sigma^2;a,b)
= \frac{\sqrt{r}}{\sigma \sqrt{2 \pi}} \frac{b^a}{\Gamma(a)}\left(\frac{1}{\sigma^2}\right)^{a+1} \exp \left(-\frac{2 b+r(\mu-\mu_0)^2}{2 \sigma^2}\right),$$
where $f_N$ and $f_{\Gamma^{-1}}$ are the PDFs of the normal distribution and the inverse-gamma distribution.
Note that the above is in the form of prior distribution for $\mu , \sigma$.

    \item The PDF of  a log-normal distribution with mean $\mu$ and standard deviation $\sigma$ is given by 
\begin{align}
\label{align:lognormal}
    f_{LN}(x;\mu, \sigma) = \frac{1}{x\sqrt{2\pi}\sigma}\exp\left[-\frac{1}{2\sigma^2}(\ln x-\mu)^2\right],\quad x>0.
\end{align}
    \item The PDF of a generalized beta of the second kind (GB2) distribution with parameters $a,b,p,q$ is given by 
\begin{align}
    f_{GB2}(x;a,b,p,q) = \frac{\frac{|a|}{b}\left(\frac x b\right)^{ap-1}}{B(p,q)\left[1+\left(\frac x b\right)^a\right]^{p+q}},\quad x>0;
\end{align}
where $B(p,q)$ is the bivariate beta function given by
\begin{align}
    B(p,q)=\int_0^1 t^{p-1}(1-t)^{q-1}dt = \frac{\Gamma(p)\Gamma(q)}{\Gamma(p+q)}.
\end{align}

%    \item The PDF of a Weibull distribution with parameters $\theta, \beta, \mu$ is given by
%$$f_{WB}(x; \theta,\beta,\mu) = \frac\beta\theta\left(\frac{x-\mu}{\theta}\right)^{(\beta-1)}\exp\left\{-\left(\frac{x-\mu}{\theta}\right)^\beta\right\},\quad x>\mu, $$
%where $\theta$, $\beta$ and $\mu$ are the scale, shape and location parameters.

    \item The PDF of a Pareto distribution with parameters $\alpha,\sigma$ is given by
$$f_P(x;\alpha,\sigma)=\frac{\alpha\sigma^\alpha}{x^{\alpha+1}}, \quad x>\sigma,$$
where $\alpha$ and $\sigma$ are the shape and location parameters.

\end{enumerate}

\section{Stability of the sampled occurrence probabilities in Model 4LN.}\label{appendix:stability}

\renewcommand {\thetable} {\Alph{section}-\arabic{table}}

Table \ref{tab:lwwptop20_stablility} shows the stability of the sampled occurrence probabilities $\bphi_k^{[t]}$ in Model 4LN.
We list the sampled occurrence probabilities of the 10 most probable words at  $t=1000, 1200, 1400, 1600, 1800$ iteration. 
We measure the stability  by using the Euclidean distance and the KL divergence.

\setcounter{table}{0}

\begin{table*}[!htbp]
    \renewcommand\arraystretch{1.2}
    \small
    \centering
    \caption{The stability of the sampled occurrence probabilities in Model 4LN.}
    \begin{tabular}{ll|ll|ll|ll|ll}
    \hline\hline
    \multicolumn{2}{c}{\textbf{1000}} & \multicolumn{2}{c}{\textbf{1200}} & \multicolumn{2}{c}{\textbf{1400}} & \multicolumn{2}{c}{\textbf{1600}} & \multicolumn{2}{c}{\textbf{1800}} \\\hline
    \multicolumn{10}{c}{\textbf{Topic 1 (Euclidean distance = 1.191$\mathbf{\times 10^{-4}}$, KL divergence = 6.114$\mathbf{\times 10^{-5}}$)}}\\
    \hline
    fell&.04540 &back&.04540 &fell&.04540 &fell&.04537 &fell&.04536 \\
    knee&.04330 &knee&.04330 &knee&.04331 &knee&.04333 &strain&.04335 \\
    ankle&.03596 &ankle&.03596 &ankle&.03597 &ankle&.03595 &lifting&.03594 \\
    injury&.03514 &injury&.03514 &injury&.03515 &injury&.03513 &shoulder&.03515 \\
    wrist&.03403 & wrist&.03403 &wrist&.03403 &wrist&.03403 &strained&.03402 \\
    slipped&.03237 &slipped&.03237 &slipped&.03239 &slipped&.03237 &slipped&.03238 \\
    foot&.02502 &foot&.02502 &foot&.02502 &foot&.02501 &foot&.02500 \\
    bruised&.02415 &bruised&.02415 &bruised&.02415 &bruised&.02414 &bruised&.02413 \\
    sprained&.01948 &sprained&.01948 &sprained&.01949 &sprained&.01947 &sprained&.01947 \\
    hand&.01866 &hand&.01866 &hand&.01865 &hand&.01865 &hand&.018637 \\
    \hline
    \multicolumn{10}{c}{\textbf{Topic 2 (Euclidean distance = 9.4342$\mathbf{\times 10^{-5}}$, KL divergence = 5.0679$\mathbf{\times 10^{-5}}$)}}\\
    \hline
    back       &.16579 &back       &.16581 &back       &.16579 &back       &.16583 &back       &.16583 \\
    strain     &.14593 &strain     &.14593 &strain     &.14593 &strain     &.14596 &strain     &.14596 \\
    lifting    &.08248 &lifting    &.08249 &lifting    &.08248 &lifting    &.08250 &lifting    &.08250 \\
    shoulder   &.04481 &shoulder   &.04476 &shoulder   &.04475 &shoulder   &.04473 &shoulder   &.04476 \\
    strained   &.03407 &strained   &.03408 &strained   &.03407 &strained   &.03408 &strained   &.03408 \\
    neck       &.02723 &neck       &.02725 &neck       &.02726 &neck       &.02727 &neck       &.02727 \\
    pain      &.01836 &pain      &.01835 &pain      &.01835 &pain      &.01835 &pain      &.01835 \\
    injury    &.01308 &injury    &.01308 &injury    &.01308 &injury    &.01309 &injury    &.01307 \\
    fell      &.01268 &fell      &.01268 &fell      &.01268 &fell      &.01268 &fell      &.01267 \\
    slipped    &.01260 &slipped    &.01260 &slipped    &.01258 &slipped    &.01257 &slipped    &.01256 \\
    \hline
    \multicolumn{10}{c}{\textbf{Topic 3 (Euclidean distance =  1.2267$\mathbf{\times 10^{-4}}$, KL divergence = 3.6168$\mathbf{\times 10^{-5}}$)}}\\
    \hline
eye&0.1472 &eye&0.1473 &eye&0.1473 &eye&0.1473 &eye&0.1473 \\
foreign&0.1185 &foreign&0.1184 &foreign&0.1185 &foreign&0.1185 &foreign&0.1185 \\
body&0.1168 &body&0.1168 &body&0.1168 &body&0.1168 &body&0.1168  \\
grinding&0.0341 &grinding&0.0341 &grinding&0.0341 &grinding&0.0341 &grinding&0.0342  \\
metal&0.0258 &metal&0.0258 &metal&0.0258 &metal&0.0258 &metal&0.0259  \\
burn&0.0234 &burn&0.0234 &burn&0.0233 &burn&0.0233 &burn&0.0233  \\
cornea&0.0205 &cornea&0.0205 &cornea&0.0205 &cornea&0.0205 &cornea&0.0205  \\
struck&0.0199 &struck&0.0199 &struck&0.0199 &struck&0.0199 &struck&0.0199  \\
shoulder&0.0156 &shoulder&0.0156 &shoulder&0.0156 &shoulder&0.0156 &shoulder&0.0156  \\
steel&0.0134 &steel&0.0134 &steel&0.0134 &steel&0.0134 &steel&0.0134  \\
    \hline
    \multicolumn{10}{c}{\textbf{Topic 4 (Euclidean distance =  8.3963$\mathbf{\times 10^{-5}}$, KL divergence = 3.2879$\mathbf{\times 10^{-5}}$)}}\\
    \hline
finger&0.0927&finger&0.0927&finger&0.0926&finger&0.0927&finger&0.0926\\
hand&0.0716&hand&0.0716&hand&0.0716&hand&0.0716&hand&0.0716 \\
laceration&0.0648&laceration&0.0648&laceration&0.0648&laceration&0.0648&laceration&0.0648 \\
struck&0.0620&struck&0.0620&struck&0.0620&struck&0.0620&struck&0.0620 \\
lacerated&0.0525&lacerated&0.0525&lacerated&0.0525&lacerated&0.0525&lacerated&0.0525 \\
cut&0.0359&cut&0.0359&cut&0.0359&cut&0.0358&cut&0.0358 \\
knife&0.0331&knife&0.0331&knife&0.0330&knife&0.0331&knife&0.0331 \\
thumb&0.0325&thumb&0.0325&thumb&0.0325&thumb&0.0325&thumb&0.0325 \\
index&0.0306&index&0.0306&index&0.0306&index&0.0306&index&0.0306 \\
slipped&0.0190&slipped&0.0190&slipped&0.0190&slipped&0.0190&slipped&0.0190 \\
caught&0.0175&caught&0.0175&caught&0.0175&caught&0.0175&caught&0.0175 \\
    \hline\hline
    \end{tabular}
    \label{tab:lwwptop20_stablility}
\end{table*}

\newpage

\bibliography{refs}
\bibliographystyle{refs}

\end{document}